# ETRAJ.JL:
## Trajectory-Based Simulation for Strong-Field Ionization

Mingyu Zhu,[1] Hongcheng Ni,[1, 2, 3, *] and Jian Wu[1, 2, 3, 4]

[1]*State Key Laboratory of Precision Spectroscopy, East China Normal University, Shanghai 200241, China*
[2]*NYU-ECNU Joint Institute of Physics, New York University at Shanghai, Shanghai 200062, China*
[3]*Collaborative Innovation Center of Extreme Optics, Shanxi University, Taiyuan, Shanxi 030006, China*
[4]*Chongqing Key Laboratory of Precision Optics, Chongqing Institute of East China Normal University, Chongqing 401121, China*

(Dated: February 26, 2025)

The dynamics of light-matter interactions in the realm of strong-field ionization has been a focal point and has attracted widespread interest. We present the `eTraj.jl` program package, designed to implement established classical/semiclassical trajectory-based methods to determine the photoelectron momentum distribution resulting from strong-field ionization of both atoms and molecules. The program operates within a unified theoretical framework that separates the trajectory-based computation into two stages: initial-condition preparation and trajectory evolution. For initial-condition preparation, we provide several methods, including the Strong-Field Approximation with Saddle-Point Approximation (SFA-SPA), SFA-SPA with Non-adiabatic Expansion (SFA-SPANE), and the Ammosov-Delone-Krainov theory (ADK), with atomic and molecular variants, as well as the Weak-Field Asymptotic Theory (WFAT) for molecules. For trajectory evolution, available options are Classical Trajectory Monte-Carlo (CTMC), which employs purely classical electron trajectories, and the Quantum Trajectory Monte-Carlo (QTMC) and Semi-Classical Two-Step model (SCTS), which include the quantum phase during trajectory evolution. The program is a versatile, efficient, flexible, and out-of-the-box solution for trajectory-based simulations for strong-field ionization. It is designed with user-friendliness in mind and is expected to serve as a valuable and powerful tool for the community of strong-field physics.

**PROGRAM SUMMARY**
*Program title:* `eTraj.jl`
*Repository link:* https://github.com/TheStarAlight/eTraj
*Licensing provisions:* Apache License, 2.0 (Apache-2.0)
*Programming language:* Julia
*Platform:* Linux (full functionality), macOS (full functionality), Windows (molecular calculation is restricted)

*Nature of problem:* Atoms and molecules exposed in an intense laser field go through complex processes of ionization through mechanisms such as multi-photon ionization and tunneling ionization. The trajectory-based methods are powerful tools for simulating these processes, and have considerable advantages over the time-dependent Schrödinger equation (TDSE) and the strong-field approximation (SFA). However, the community lacks a unified theoretical framework for trajectory-based methods, and there are no public-available code that implements the schemes.

*Solution method:* We developed a general, efficient, flexible, and out-of-the-box solution for trajectory-based simulation program named after `eTraj.jl` using the Julia programming language. This program conducts trajectory-based classical/semiclassical simulations of photoelectron dynamics under the single-active-electron approximation and the Born-Oppenheimer approximation. It supports multiple methods, including the SFA-SPA, SFA-SPANE, ADK and WFAT for initial condition preparation. Additionally, it incorporates the CTMC, QTMC and SCTS methods for trajectory evolution. The program is written in a clear and concise manner, and features versatility, extensibility, and usability.

*Additional comments including restrictions and unusual features:* A detailed documentation is available at https://thestaralight.github.io/eTraj.jl/stable/. The package has been tested for compatibility with Julia versions 1.9 to 1.11 and is expected to remain compatible with newer Julia versions released after the test date.

* hcni@lps.ecnu.edu.cn

**CONTENTS**

## I. INTRODUCTION

The interaction between light and matter has been a subject of widespread investigation since the inception of quantum mechanics. The development of laser technology has led to remarkable advances in both light intensity and spectroscopic precision, enabling unprecedented exploration of light-matter interactions under extreme conditions [1–3].

When laser intensities exceed TW/cm$^2$, the light-matter interaction enters a non-perturbative regime where conventional perturbation theory becomes inadequate, giving rise to various novel strong-field phenomena such as the above-threshold ionization (ATI) [4–6], tunneling ionization [7–11], high-harmonic generation (HHG) [12–17] and non-sequential double ionization (NSDI) [18, 19]. Theoretical investigations of these non-perturbative phenomena have progressed substantially over recent decades. The most rigorous approach involves numerical solution of the time-dependent Schrödinger equation (TDSE) [20–24]. While this method provides highly accurate results, its computational complexity restricts applications primarily to few-dimensional systems. Furthermore, the abstract nature of TDSE calculations often obscures the underlying physical mechanisms. An alternative approach, the strong-field approximation (SFA) [7, 25–27], has been widely adopted based on two key assumptions: first, that the initial state remains unperturbed by the laser field until ionization occurs; and second, that the photoelectron's post-ionization dynamics proceed without influence from the binding potential (effectively treating it as short-range). These approximations enable analytical treatment of the problem, thereby providing valuable physical insights into the underlying mechanisms. Nevertheless, the SFA framework exhibits limitations, particularly in scenarios where Coulomb interactions play a significant role, potentially leading to qualitative discrepancies with experimental observations [28–35].

A substitute strategy to overcome these limitations is the *Classical-Trajectory Monte-Carlo* (CTMC) method [36, 37], which employs an ensemble of classical electrons evolving under combined laser and Coulomb fields. This methodology has been extended to incorporate tunneling ionization effects by initializing electron trajectories at the tunnel exit coordinate [6, 15, 38]. The photoelectron momentum distribution (PMD) is subsequently obtained through statistical analysis of these classical trajectories. Although the CTMC approach is fundamentally classical in nature, quantum mechanical effects can be effectively incorporated through the introduction of trajectory-dependent phases. Examples include the *Trajectory-based Coulomb-SFA* (TC-SFA) [39, 40], the *Quantum-Trajectory Monte Carlo* (QTMC) [41–44], and the *Semiclassical Two-Step Model* (SCTS) [45–50]. Another approach, the *Coulomb Quantum-orbit SFA* (CQSFA) [51–56], addresses the inverse problem by identifying all trajectories that result in the same final momenta. These trajectory-based semiclassical methods offer notable advantages over the TDSE and direct SFA methods due to their lower demand on computational resources, as well as the clarity they provide in understanding the physical picture.

After years of development, various trajectory-based classical/semiclassical methods have emerged, yet a unified theoretical framework remains to be established. In addition, developing a library that not only implements existing methods but also in a way that is both computationally efficient and easy to maintain can significantly enhance research in strong-field ionization. To meet these challenges, we introduce `eTraj.jl`, a program package written in Julia [57]. Julia was chosen for its extraordinary balance of performance, ease of use, and simplicity in deployment, which are crucial for scientific computing: it combines Python-like syntax with C-like speed due to its just-in-time (JIT) compilation; offers a user-friendly syntax that simplifies coding and enhances productivity, enabling researchers to focus on the working problem; includes built-in support for parallel computing, allowing efficient multicore utilization without complex setup; what's more, programs written in Julia are easy to deploy across different environments, ensuring accessibility and broad applicability. `eTraj` leverages these features to provide an efficient, versatile, flexible, and out-of-the-box solution for classical/semiclassical trajectory simulations, advancing research in strong-field ionization.

The remainder of this article is organized as follows. Section II presents the theoretical framework of `eTraj`, with particular emphasis on the initial conditions and phase methods employed in trajectory-based approaches. Section III provides a comprehensive description of the `eTraj` architecture and implementation. Section IV demonstrates the capability and versatility of `eTraj` through several representative applications. Section V summarizes the article and discusses future prospects.

Atomic units (a.u.) are used in this article unless stated otherwise.

## II. THEORETICAL FRAMEWORK

### A. Initial Conditions

Multiple theoretical frameworks for strong-field ionization provide methodologies to determine the initial conditions of classical electrons in trajectory simulations. These initial conditions are characterized by three essential properties:

- Initial position $\boldsymbol{r_0}$, i.e., the tunneling exit position;

- Initial momentum $\boldsymbol{k_0}$ [58];

- The corresponding ionization probability $W$ carried by each electron sample, which depends on the time-dependent laser field and properties of the target atom/molecule.

In this section we briefly revisit the available theories [3] implemented in `eTraj`.

#### 1. Strong-Field Approximation with Saddle-Point Approximation (SFA-SPA)

The *Strong-Field Approximation (SFA)* is originated from the Keldysh theory of strong-field ionization [7, 25–27]. In contrast to perturbative methods and adiabatic tunneling theories, the SFA framework encompasses both multi-photon and tunneling processes in laser-atom interactions, as it comprehensively incorporates non-adiabatic effects of the laser-atom coupling. This comprehensive applicability has facilitated the extensive use of SFA in theoretical studies of strong-field ionization.

We consider an electron evolving under a combined field of the Coulomb field $V(\boldsymbol{r})$ of the parent ion and the laser field $\boldsymbol{F}(t) = -\partial_t \boldsymbol{A}(t)$, where $\boldsymbol{F}(t)$ and $\boldsymbol{A}(t)$ are the electric field and vector potential of the laser field, respectively. Under the length gauge (LG), its Hamiltonian reads

$$H^{\mathrm{LG}} = \frac{1}{2}\boldsymbol{p}^2 + V(\boldsymbol{r}) + \boldsymbol{F}(t) \cdot \boldsymbol{r}. \tag{1}$$

Denoting $|\Psi_0\rangle = |\psi_0\rangle \mathrm{e}^{\mathrm{i}I_p t}$ as the unperturbed initial state with ionization potential of $I_p$, $|\Psi_{\boldsymbol{p}}\rangle$ as the continuum state of momentum $\boldsymbol{p}$, and

$$U(t_\mathrm{f}, t_0) = \exp\left[-\mathrm{i}\int_{t_0}^{t_\mathrm{f}} H^{\mathrm{LG}}(\tau)\mathrm{d}\tau\right] \tag{2}$$

the time-evolution operator, the transition amplitude between the initial state (at $t_0$) and the final state of momentum $\boldsymbol{p}$ (at $t_\mathrm{f}$) is written as

$$M_{\boldsymbol{p}} = \langle \Psi_{\boldsymbol{p}}|U(t_\mathrm{f}, t_0)|\Psi_0\rangle. \tag{3}$$

Here lies the key idea of SFA: when the influence of the Coulomb field on the ionized electrons is weak compared with that of the external laser field, we may neglect the influence of the Coulomb field in the expression of $M_{\boldsymbol{p}}$ by replacing the time-evolution operator with a Coulomb-free one $U_\mathrm{f}$, and meanwhile replacing the continuum state with the Volkov state $|\Psi_{\boldsymbol{p}}^{\mathrm{V}}\rangle$ which represents a free electron evolving under the same laser field:

$$M_{\boldsymbol{p}} \approx \langle \Psi_{\boldsymbol{p}}^{\mathrm{V}}|U_\mathrm{f}(t_\mathrm{f}, t_0)|\Psi_0\rangle, \tag{4}$$

where the Volkov state under the LG is the product of a plane wave and a phase factor:

$$|\Psi_{\boldsymbol{p}}^{\mathrm{V}}\rangle = |\boldsymbol{p} + \boldsymbol{A}(t)\rangle \exp\left\{-\mathrm{i}\int^t \frac{1}{2}[\boldsymbol{p} + \boldsymbol{A}(\tau)]^2\mathrm{d}\tau\right\}. \tag{5}$$

Consequently, the expression for $M_{\boldsymbol{p}}$ becomes

$$M_{\boldsymbol{p}} = -\mathrm{i} \int_{t_0}^{t_f} \langle \boldsymbol{p} + \boldsymbol{A}(\tau) | \boldsymbol{F}(\tau) \cdot \boldsymbol{r} | \psi_0 \rangle \, \mathrm{e}^{-\mathrm{i}S_{\boldsymbol{p}}(\tau)} \mathrm{d}\tau, \tag{6}$$

and we note that here we have extracted the phase factor of $|\Psi_0\rangle$ and combined it with that of the Volkov state $|\Psi_{\boldsymbol{p}}^{\mathrm{V}}\rangle$, giving the phase

$$S_{\boldsymbol{p}}(t) = -\int^t \left\{ \frac{1}{2}[\boldsymbol{p} + \boldsymbol{A}(\tau)]^2 + I_{\mathrm{p}} \right\} \mathrm{d}\tau. \tag{7}$$

Inserting

$$\frac{\partial}{\partial t} \langle \boldsymbol{p} + \boldsymbol{A}(t) | = \mathrm{i} \langle \boldsymbol{p} + \boldsymbol{A}(t) | \left[ \boldsymbol{F}(t) \cdot \boldsymbol{r} \right] \tag{8}$$

into the above expression of $M_{\boldsymbol{p}}$ [Eq. (6)], after integration by parts, one obtains

$$\begin{aligned} M_{\boldsymbol{p}} &= -\int_{t_0}^{t_f} \mathrm{d}\tau \frac{\partial}{\partial \tau} [\langle \boldsymbol{p} + \boldsymbol{A}(\tau) | \psi_0 \rangle] \mathrm{e}^{-\mathrm{i}S_{\boldsymbol{p}}(\tau)} \\ &= -\langle \boldsymbol{p} + \boldsymbol{A}(\tau) | \psi_0 \rangle \, \mathrm{e}^{-\mathrm{i}S_{\boldsymbol{p}}(\tau)} \Big|_{t_0}^{t_f} + \int_{t_0}^{t_f} \mathrm{d}\tau \, \langle \boldsymbol{p} + \boldsymbol{A}(\tau) | \psi_0 \rangle \cdot [-\mathrm{i}S_{\boldsymbol{p}}'(\tau)] \mathrm{e}^{-\mathrm{i}S_{\boldsymbol{p}}(\tau)}. \end{aligned} \tag{9}$$

An additional saddle-point approximation (SPA) facilitates preparation of initial conditions of the electron trajectories. Given that the variation of the phase factor $\mathrm{e}^{\mathrm{i}S_{\boldsymbol{p}}(t)}$ is significantly more sensitive to changes in $t$ compared to the prefactor, the integrand in Eq. (9) tends to oscillate rapidly in its complex phase. Consequently, contributions tend to cancel out over most of the domain, except near points where the phase $S_{\boldsymbol{p}}(t)$ stabilizes, specifically at the saddle points. The saddle points $t_{\mathrm{s}} = t_{\mathrm{r}} + \mathrm{i}t_{\mathrm{i}}$ are the zeroes of the derivative of the complex function $S_{\boldsymbol{p}}(t)$, which satisfy

$$-S_{\boldsymbol{p}}'(t_{\mathrm{s}}) = \frac{1}{2}[\boldsymbol{p} + \boldsymbol{A}(t_{\mathrm{s}})]^2 + I_{\mathrm{p}} = 0. \tag{10}$$

The integral, representing the second term on the right-hand side of Eq. (9), contributes significantly only in the neighborhood of the endpoints $t_0$, $t_f$, and the saddle points $t_{\mathrm{s}}$. The contributions near the endpoints largely cancel out the first term. Therefore, the $M_{\boldsymbol{p}}$ is now approximated with the integration around the saddle points:

$$M_{\boldsymbol{p}} \approx \sum_{t_{\mathrm{s}}} \int_{C_{t_{\mathrm{s}}}} \mathrm{d}\tau \, \langle \boldsymbol{p} + \boldsymbol{A}(\tau) | \psi_0 \rangle \cdot [-\mathrm{i}S_{\boldsymbol{p}}'(\tau)] \mathrm{e}^{-\mathrm{i}S_{\boldsymbol{p}}(\tau)}, \tag{11}$$

with $C_{t_{\mathrm{s}}}$ the integration contour following the steepest-descent path related to $t_{\mathrm{s}}$.

Further evaluation of the prefactor $\tilde{\psi}_0(\boldsymbol{k})|_{\boldsymbol{k} = \boldsymbol{p} + \boldsymbol{A}(t)} = \langle \boldsymbol{p} + \boldsymbol{A}(t) | \psi_0 \rangle$ (i.e., the momentum-space wavefunction) in the vicinity of the saddle points in Eq. (11) is essential before applying the SPA. We assume the field points towards the $+z$ axis, for an atom target at the $(l, m)$ state with ionization potential $I_{\mathrm{p}}$, its wavefunction behaves asymptotically as [59]

$$\psi_0(\boldsymbol{r}) \sim 2C_{\kappa l}\kappa^{3/2}(\kappa r)^{n^*-1}\mathrm{e}^{-\kappa r}Y_{lm}(\hat{\boldsymbol{r}}) \tag{12}$$

for $\kappa r \gg 1$, with $\kappa = \sqrt{2I_{\mathrm{p}}}$, $n^* = Z/\kappa$ the effective principal quantum number, $Z$ the charge of the residual ion, $Y_{lm}$ the spherical harmonics, and $C_{\kappa l}$ the asymptotic coefficient for atoms, which can be approximated using the Hartree approximation formula [60]

$$C_{\kappa l}^2 = \frac{2^{2n^*-2}}{n^*(n^*+l)!(n^*-l-1)!}. \tag{13}$$

For atomic hydrogen at the ground state we have $C_{\kappa l} = 1$. Moreover, for non-integer $n^*$, the formula can be naturally extended by replacing the factorials $x!$ with Gamma functions $\Gamma(x+1)$, i.e.,

$$C_{\kappa l}^2 = \frac{2^{2n^*-2}}{n^*\Gamma(n^*+l+1)\Gamma(n^*-l)}. \tag{14}$$

Near the saddle points, corresponding to the limit where $k^2 \to -\kappa^2$, the form of $\tilde{\psi}_0(\boldsymbol{k})$ is dictated by the asymptotic behavior of the wavefunction [61]:

$$\tilde{\psi}_0(\boldsymbol{k}) = \frac{C_{\kappa l}}{\sqrt{\pi}} \frac{2^{n^*+3/2}\kappa^{2n^*+1/2}\Gamma(n^*+1)}{(k^2+\kappa^2)^{n^*+1}} Y_{lm}(\hat{\boldsymbol{k}}). \tag{15}$$

By substituting the aforementioned expression into Eq. (11) and using the definition of $S_{\boldsymbol{p}}(t)$ given in Eq. (7), we derive

$$M_{\boldsymbol{p}} = \mathrm{i}\frac{C_{\kappa l}}{\sqrt{\pi}} 2^{1/2}\kappa^{2n^*+1/2}\Gamma(n^*+1) \sum_{t_s} \int_{C_{t_s}} \frac{Y_{lm}[\hat{\boldsymbol{k}}(\tau)]}{[S'_{\boldsymbol{p}}(\tau)]^{n^*}} \mathrm{e}^{\mathrm{i}S_{\boldsymbol{p}}(\tau)}\mathrm{d}\tau, \tag{16}$$

where $\hat{\boldsymbol{k}}(\tau)$ denotes the complex unit vector along $\boldsymbol{k}(\tau) = \boldsymbol{p} + \boldsymbol{A}(\tau)$. The evaluation method for spherical harmonics with complex arguments follows Appendix B of Ref. [62]; see also the note [63].

To address cases where the integrand exhibits a singularity at $t_s$, a modified SPA approach can be applied (see Appendix B of Ref. [64]):

$$\int_{C_{t_s}} \frac{Y_{lm}[\hat{\boldsymbol{k}}(\tau)]}{[S'_{\boldsymbol{p}}(\tau)]^{n^*}} \mathrm{e}^{\mathrm{i}S_{\boldsymbol{p}}(\tau)}\mathrm{d}\tau \approx \frac{Y_{lm}[\hat{\boldsymbol{k}}(t_s)]}{[S''_{\boldsymbol{p}}(t_s)]^{n^*}} \int_{C_{t_s}} \frac{\mathrm{e}^{\mathrm{i}S_{\boldsymbol{p}}(\tau)}}{(\tau-t_s)^{n^*}}\mathrm{d}\tau$$
$$\approx \frac{Y_{lm}[\hat{\boldsymbol{k}}(t_s)]}{[S''_{\boldsymbol{p}}(t_s)]^{n^*}} \cdot \mathrm{i}^{n^*} \frac{\Gamma(n^*/2)}{2\Gamma(n^*)} \sqrt{\frac{2\pi}{-\mathrm{i}S''_{\boldsymbol{p}}(t_s)}} [-2\mathrm{i}S''_{\boldsymbol{p}}(t_s)]^{n^*/2} \mathrm{e}^{\mathrm{i}S_{\boldsymbol{p}}(t_s)}. \tag{17}$$

Through this approach, we arrive at the following expression for the transition amplitude:

$$M_{\boldsymbol{p}} = c_{n^*} C_{\kappa l} \sum_{t_s} \frac{Y_{lm}[\hat{\boldsymbol{k}}(t_s)]}{[S''_{\boldsymbol{p}}(t_s)]^{(n^*+1)/2}} \mathrm{e}^{-\mathrm{i}S_{\boldsymbol{p}}(t_s)}, \tag{18}$$

where $c_{n^*} = \mathrm{i}^{(n^*-5)/2}2^{n^*/2+1}\kappa^{2n^*+1/2}\Gamma(n^*/2+1)$ represents a constant coefficient.

The SFA phase $S_{\boldsymbol{p}}(t_s)$ is determined by evaluating the integral

$$S_{\boldsymbol{p}}(t_s) = -\int_{t_s}^{\infty}\mathrm{d}\tau\left\{\frac{1}{2}[\boldsymbol{p}+\boldsymbol{A}(\tau)]^2 + I_p\right\}$$
$$= \left(-\int_{t_s}^{t_r} - \int_{t_r}^{\infty}\right)\mathrm{d}\tau\left\{\frac{1}{2}[\boldsymbol{p}+\boldsymbol{A}(\tau)]^2 + I_p\right\} \tag{19}$$
$$= S_{\boldsymbol{p},\mathrm{tun}} + S_{\boldsymbol{p},\mathrm{traj}},$$

where $S_{\boldsymbol{p},\mathrm{tun}}$ and $S_{\boldsymbol{p},\mathrm{traj}}$ denote the complex phases accumulated during the tunneling process and the subsequent motion in the continuum, respectively. The phase $S_{\boldsymbol{p},\mathrm{tun}}$ corresponds to an imaginary time interval (from $t_s$ to $t_r$), during which the electron traverses the potential barrier with an "imaginary" momentum; its real part signifies the quantum phase, whereas its imaginary part is associated with the ionization probability.

To apply the SFA for preparing initial conditions of photoelectrons, we assume that the electron is emitted at time $t_r$ from the tunnel exit $\boldsymbol{r}_0$ with momentum $\boldsymbol{k}_0 = \boldsymbol{k}(t_r)$. The initial momentum $\boldsymbol{k}_0$, when the Coulomb interaction with the nucleus is neglected, is correlated with the final momentum $\boldsymbol{p}$ via the relationship

$$\boldsymbol{p} = \boldsymbol{k}_0 - \int_{t_r}^{\infty}\boldsymbol{F}(\tau)\mathrm{d}\tau = \boldsymbol{k}_0 - \boldsymbol{A}(t_r). \tag{20}$$

The initial position $\boldsymbol{r}_0$, corresponding to the tunnel exit, is determined by constructing a quantum tunneling trajectory. This trajectory starts at a point with vanishing real part, representing the tunnel entrance; the electron tunnels through the barrier over the time interval $t_s$ to $t_r$ and emerges as a classical electron at the tunnel exit $\boldsymbol{r}_0$ with real position and momentum. Thus, the expression for the initial position becomes:

$$\boldsymbol{r}_0^{\mathrm{SFA-SPA}} = \Re\int_{t_s}^{t_r}[\boldsymbol{p}+\boldsymbol{A}(\tau)]\mathrm{d}\tau = \Im\int_0^{t_i}\boldsymbol{A}(t_r+\mathrm{i}\tau)\mathrm{d}\tau, \tag{21}$$

with $\Re$ and $\Im$ denoting the real and imaginary parts, respectively. The probability density of the electron sample in the final momentum space $\boldsymbol{p}$ is given by

$$\mathrm{d}W^{\mathrm{SFA-SPA}}/\mathrm{d}\boldsymbol{p} = \sum_{t_s} \left|\mathcal{P}_{\boldsymbol{p}}^{\mathrm{SFA-SPA}}(t_s)\right|^2 \exp\{-2\Im S_{\boldsymbol{p},\mathrm{tun}}(t_s)\}, \tag{22}$$

where the prefactor encompasses all coefficients:

$$\mathcal{P}_{\boldsymbol{p}}^{\mathrm{SFA-SPA}}(t_s) = c_{n^*} \frac{C_{\kappa l} Y_{lm}[\hat{\boldsymbol{k}}(t_s)]}{[S_{\boldsymbol{p}}''(t_s)]^{(n^*+1)/2}} = c_{n^*} \frac{C_{\kappa l} Y_{lm}[\hat{\boldsymbol{k}}(t_s)]}{\{[\boldsymbol{p} + \boldsymbol{A}(t_s)] \cdot \boldsymbol{F}(t_s)\}^{(n^*+1)/2}}. \tag{23}$$

It should be noted that the ionization probability in Eq. (22) is formulated in terms of the final momentum coordinates $\boldsymbol{p} = (p_x, p_y, p_z)$. However, in trajectory simulations, initial electrons are sampled in the $(t_r, \boldsymbol{k}_t)$ coordinate system, with $\boldsymbol{k}_t$ being the initial transverse momentum. Consequently, if we sample the initial electrons within such a coordinate system, a Jacobian must be introduced as a prefix to the ionization probability. Assuming laser propagation along the $z$ axis and polarization in the $xy$ plane, the transformed expression reads

$$\mathrm{d}W^{\mathrm{SFA-SPA}}/\mathrm{d}t_r \mathrm{d}\boldsymbol{k}_t = \sum_{t_s} J(t_r, k_\perp)\left|\mathcal{P}_{\boldsymbol{p}}^{\mathrm{SFA-SPA}}(t_s)\right|^2 \exp\left(-2\Im S_{\boldsymbol{p},\mathrm{tun}}(t_s)\right), \tag{24}$$

where $k_\perp$ represents the projection of $\boldsymbol{k}_t$ onto the polarization plane (i.e., the $xy$ plane), and the Jacobian is defined as

$$J(t_r, k_\perp) = \left|\frac{\partial(p_x, p_y)}{\partial(t_r, k_\perp)}\right| = \left|\begin{matrix} \partial p_x/\partial t_r & \partial p_x/\partial k_\perp \\ \partial p_y/\partial t_r & \partial p_y/\partial k_\perp \end{matrix}\right|. \tag{25}$$

### 2. SFA-SPA with Non-adiabatic Expansion (SFA-SPANE)

When the Keldysh parameter $\gamma = \omega\kappa/F_0$ is small (where $\omega$ denotes the laser angular frequency and $F_0$ represents the peak field strength), non-adiabatic effects become less significant. Under these conditions, a non-adiabatic expansion scheme can be implemented to develop an approximation based on the SFA-SPA, termed the *SFA-SPA with Non-adiabatic Expansion (SFA-SPANE)* [3, 65–69]. It captures non-adiabatic effects to a considerable degree and yields results that are comparable to those obtained from the SFA-SPA for relatively small values of the Keldysh parameter. SFA-SPANE comes with a closed analytical form, avoiding the necessity to solve the saddle-point equation, thereby speeding up the calculation.

The SFA-SPANE method is applicable when the Keldysh parameter is small, and the non-adiabatic effect is insignificant, which corresponds to the small-$t_i$ case. We expand the vector potential $\boldsymbol{A}(t_s) = \boldsymbol{A}(t_r + \mathrm{i}t_i)$ in the SFA-SPA around $t_i = 0$, up to the second order of $t_i$:

$$\boldsymbol{A}(t_r + \mathrm{i}t_i) = \boldsymbol{A}(t_r) - \mathrm{i}t_i\boldsymbol{F}(t_r) + \frac{1}{2}t_i^2\boldsymbol{F}'(t_r) + o(t_i^2). \tag{26}$$

Inserting Eq. (26) into the saddle-point equation in the SFA-SPA [Eq. (10)] leads to

$$\boldsymbol{k}(t_r) \cdot \boldsymbol{F}(t_r) \approx 0 \tag{27}$$

and

$$t_i \approx \sqrt{\frac{k^2(t_r) + \kappa^2}{F^2(t_r) - \boldsymbol{k}(t_r) \cdot \boldsymbol{F}'(t_r)}}, \tag{28}$$

which allow for the derivation of analytical expressions of the ionization probability and other quantities.

The initial position $\boldsymbol{r}_0$ in SFA-SPANE is given by

$$\boldsymbol{r}_0^{\mathrm{SFA-SPANE}} = \Im \int_0^{t_i} \boldsymbol{A}(t_r + \mathrm{i}\tau)\mathrm{d}\tau = -\frac{\boldsymbol{F}}{2}\frac{k_t^2 + \kappa^2}{F^2 - \boldsymbol{k}_0 \cdot \boldsymbol{F}'}. \tag{29}$$

The $\Im S_{\boldsymbol{p},\mathrm{tun}}$ term, which is related to the ionization probability, within the context of SFA-SPANE, is expressed as

$$
\begin{aligned}
\Im S_{\boldsymbol{p},\mathrm{tun}} &\approx \Im \int_{t_{\mathrm{r}}}^{t_{\mathrm{s}}} \mathrm{d}\tau \left\{ \frac{1}{2} \left[ \boldsymbol{p} + \boldsymbol{A}(t_{\mathrm{r}}) - \mathrm{i}t_{\mathrm{i}}\boldsymbol{F}(t_{\mathrm{r}}) + \frac{1}{2}t_{\mathrm{i}}^2 \boldsymbol{F}'(t_{\mathrm{r}}) \right]^2 + I_{\mathrm{p}} \right\} \\
&\approx \left[ I_{\mathrm{p}} + \frac{1}{2}k^2(t_{\mathrm{r}}) \right] t_{\mathrm{i}} - [F^2(t_{\mathrm{r}}) - \boldsymbol{k}(t_{\mathrm{r}}) \cdot \boldsymbol{F}'(t_{\mathrm{r}})] \frac{t_{\mathrm{i}}^3}{6} \\
&= \frac{1}{3} \frac{(k^2 + \kappa^2)^{3/2}}{\sqrt{F^2 - \boldsymbol{k}_0 \cdot \boldsymbol{F}'}}.
\end{aligned}
\tag{30}
$$

Then follows the ionization probability,

$$
\mathrm{d}W^{\mathrm{SFA-SPANE}}/\mathrm{d}\boldsymbol{p} = \left| \mathscr{P}_{\boldsymbol{p}}^{\mathrm{SFA-SPANE}}(t_{\mathrm{s}}) \right|^2 \exp\left[ -\frac{2}{3} \frac{(k_{\mathrm{t}}^2 + \kappa^2)^{3/2}}{\sqrt{F^2 - \boldsymbol{k}_0 \cdot \boldsymbol{F}'}} \right],
\tag{31}
$$

where $\boldsymbol{k}_{\mathrm{t}}$ corresponds to $\boldsymbol{k}(t_{\mathrm{r}})$ within the framework of SFA-SPANE due to the vanishing initial longitudinal momentum, as indicated by Eq. (27). The prefactor reads

$$
\mathscr{P}_{\boldsymbol{p}}^{\mathrm{SFA-SPANE}}(t_{\mathrm{s}}) = c_{n^*} \frac{C_{\kappa l} Y_{lm}[\hat{\boldsymbol{k}}(t_{\mathrm{s}})]}{\left[ (k_{\mathrm{t}}^2 + \kappa^2)(F^2 - \boldsymbol{k}_0 \cdot \boldsymbol{F}') \right]^{(n^*+1)/4}}.
\tag{32}
$$

### 3. Ammosov-Delone-Krainov (ADK)

The *Ammosov-Delone-Krainov (ADK)* theory provides a framework for investigating adiabatic tunneling in strong-field ionization and represents, essentially, the adiabatic limit of the SFA.

In the adiabatic limit, the laser field can be considered static; consequently, we have $\boldsymbol{F}'(t) = 0$, with higher-order derivatives of $\boldsymbol{F}(t)$ also remaining zero. Substituting it into the ionization probability of SFA-SPANE [Eq. (31)] gives

$$
\mathrm{d}W^{\mathrm{ADK}}/\mathrm{d}\boldsymbol{p} = \left| \mathscr{P}_{\boldsymbol{p}}^{\mathrm{ADK}}(t_{\mathrm{s}}) \right|^2 \exp\left[ -\frac{2}{3} \frac{(k_{\mathrm{t}}^2 + \kappa^2)^{3/2}}{F} \right],
\tag{33}
$$

where the prefactor reads

$$
\mathscr{P}_{\boldsymbol{p}}^{\mathrm{ADK}}(t_{\mathrm{s}}) = c_{n^*} \frac{C_{\kappa l} Y_{lm}[\hat{\boldsymbol{k}}(t_{\mathrm{s}})]}{\left[ (k_{\mathrm{t}}^2 + \kappa^2)F^2 \right]^{(n^*+1)/4}},
\tag{34}
$$

with $t_{\mathrm{i}} = \sqrt{k_{\mathrm{t}}^2 + \kappa^2}/F$. Expanding Eq. (33) under the small-$k_{\mathrm{t}}$ approximation results in

$$
\mathrm{d}W^{\mathrm{ADK}}/\mathrm{d}\boldsymbol{p} \propto \exp\left( -\frac{2\kappa^3}{3F} \right) \exp\left( -\frac{\kappa k_{\mathrm{t}}^2}{F} \right),
\tag{35}
$$

which is actually the exponential term of the well-known ADK rate. It should be noted, however, that the outcome of our approach—specifically, the application of the adiabatic limit within the SFA-SPA framework—differs slightly from the actual ADK rate in terms of the prefactor. This is because the SFA framework neglects Coulomb potential in the final state, which has been shown to result in a lower ionization rate. To address this discrepancy, an additional Coulomb-correction (CC) factor [Eq. (A15)], introduced into Eq. (33), helps bridge the gap:

$$
C^{\mathrm{CC}} = \left( \frac{2\kappa^3}{F} \right)^{n^*} (1 + 2\gamma/e)^{-2n^*} \left[ \Gamma\left( \frac{n^*}{2} + 1 \right) \right]^{-2}.
\tag{36}
$$

We note that this CC factor is implemented in all initial-condition methods that are derived from the SFA. For more details, we refer the readers to Appendix A.

The tunnel exit is determined using the same methodology:

$$\boldsymbol{r}_0^{\text{ADK}} = \Im \int_0^{t_i} \boldsymbol{A}(t_r + i\tau) d\tau = -\frac{\boldsymbol{F}}{2} \frac{k_t^2 + \kappa^2}{F^2}, \tag{37}$$

a result that we associate with the "$I_p/F$" model. However, there is a subtle distinction in that we replace the ionization potential $I_p = \kappa^2/2$ with the effective potential $\tilde{I}_p = (\kappa^2 + k_t^2)/2$ to account for the initial kinetic energy, ensuring that adiabatic tunneling is accurately described by the condition $E = k_t^2/2 + \boldsymbol{r}_0^{\text{ADK}} \cdot \boldsymbol{F} = -I_p$.

#### 4. Molecular SFA-SPA/SFA-SPANE/ADK

The atomic SFA theory and its adiabatic variants discussed in Sections II A 1 through II A 3 can be systematically extended to molecular systems [70–73]. Under the Born-Oppenheimer approximation [74] and the single-active-electron (SAE) approximation, the strong-field ionization of molecules can be modeled as the interaction between the laser field and the ionizing orbital [often the highest occupied molecular orbital (HOMO)] $\psi_0(\boldsymbol{r})$ within the effective potential of the parent ion.

To generalize the atomic SFA to the molecular SFA (MO-SFA), we start from the transition amplitude given by Eq. (11). In the molecular frame (MF), the asymptotic wavefunction can be expanded into spherical harmonics:

$$\psi_0^{\text{MF}}(\boldsymbol{r}) \sim \sum_{l,m} 2C_{lm} \kappa^{3/2} (\kappa r)^{n^*-1} e^{-\kappa r} Y_{lm}(\hat{\boldsymbol{r}}), \tag{38}$$

where $C_{lm}$ are asymptotic coefficients, and we continue to use $n^* = Z/\kappa$ for simplicity, although it no longer represents the effective principal quantum number. We assume that in the field frame (FF), the field $\boldsymbol{F}$ points along the $z$ axis, and the rotation $\hat{\boldsymbol{R}}$ from the FF to the MF can be defined via a set of Euler angles $(\phi, \theta, \chi)$ following the $z - y' - z''$ convention, which satisfies

$$\psi_0^{\text{MF}}(\hat{\boldsymbol{R}}\boldsymbol{r}) = \psi_0^{\text{FF}}(\boldsymbol{r}). \tag{39}$$

Utilizing the Wigner-$D$ matrix, the rotated spherical harmonic function can be expressed as a linear combination of spherical harmonics of the same order $l$:

$$\hat{\boldsymbol{R}}(\phi, \theta, \chi) Y_{lm} = \sum_{m'} D_{m'm}^l(\phi, \theta, \chi) Y_{lm'}, \tag{40}$$

and the asymptotic behavior of the wavefunction in the FF is obtained by substituting Eq. (40) into Eq. (38), yielding

$$\psi_0^{\text{FF}}(\boldsymbol{r}) \sim \sum_{l,m,m'} 2C_{lm} D_{m'm}^l(\phi, \theta, \chi) \kappa^{3/2} (\kappa r)^{n^*-1} e^{-\kappa r} Y_{lm'}(\hat{\boldsymbol{r}}). \tag{41}$$

It becomes evident that the primary distinction between the molecular and atomic versions of the theory lies in the formulation of the prefactor $\mathcal{P}_{\boldsymbol{p}}(t_s)$; the expressions for the tunneling exit position and the initial momentum remain unchanged. Following the same procedure outlined in Secs. II A 1 to II A 3, we derive the prefactor $\mathcal{P}_{\boldsymbol{p}}$ applicable to molecules:

$$\mathcal{P}_{\boldsymbol{p}}^{\text{SFA-SPA}}(t_s) = c_{n^*} \frac{\sum_{l,m,m'} C_{lm} D_{m'm}^l(\phi, \theta, \chi) Y_{lm'}[\hat{\boldsymbol{k}}(t_s)]}{\{[\boldsymbol{p} + \boldsymbol{A}(t_s)] \cdot \boldsymbol{F}(t_s)\}^{(n^*+1)/2}}, \tag{42}$$

$$\mathcal{P}_{\boldsymbol{p}}^{\text{SFA-SPANE}}(t_s) = c_{n^*} \frac{\sum_{l,m,m'} C_{lm} D_{m'm}^l(\phi, \theta, \chi) Y_{lm'}[\hat{\boldsymbol{k}}(t_s)]}{\left[(k_t^2 + \kappa^2)(F^2 - \boldsymbol{k}_0 \cdot \boldsymbol{F'})\right]^{(n^*+1)/4}}, \tag{43}$$

$$\mathcal{P}_{\boldsymbol{p}}^{\text{ADK}}(t_s) = c_{n^*} \frac{\sum_{l,m,m'} C_{lm} D_{m'm}^l(\phi, \theta, \chi) Y_{lm'}[\hat{\boldsymbol{k}}(t_s)]}{\left[(k_t^2 + \kappa^2)F^2\right]^{(n^*+1)/4}}. \tag{44}$$

Furthermore, it should be noted that upon introducing an additional Coulomb-correction factor [Eq. (36)], the ionization rate conforms to the original MO-ADK theory [71] within the adiabatic and small-$k_t$ limits.

## 5. Weak-Field Asymptotic Theory (WFAT)

The *Weak-Field Asymptotic Theory (WFAT)* generalizes the tunneling ionization from isotropic atomic potentials to arbitrary molecular potentials [75–83]. Compared with the MO-ADK theory, the WFAT naturally accounts for the influence of the permanent dipole moment of the molecule, and can, in its integral representation, calculate the structure factors [a similar concept to the asymptotic coefficients $C_{lm}$ in Eq. (38)] based on the wavefunction close to the core, rather than using the wavefunction in the asymptotic region, allowing for enhanced accuracy in numerical simulations.

The formulation of the WFAT is based on the expansion in the parabolic coordinates. The total ionization rate $w$, is split into different parabolic channels:

$$w^{\text{WFAT}} = \sum_{\nu} w_{\nu}, \tag{45}$$

where $w_{\nu}$ are partial rates associated with parabolic quantum number indices $\nu = (n_{\xi}, m)$, with $n_{\xi} = 0, 1, 2, \cdots$ and $m = 0, \pm 1, \pm 2, \cdots$. In the leading-order approximation of the WFAT, the partial rates can be separated into two factors, namely the structural part $|G_{\nu}(\theta, \chi)|^2$ and the field part $\mathcal{W}_{\nu}(F)$:

$$w_{\nu} = |G_{\nu}(\theta, \chi)|^2 \, \mathcal{W}_{\nu}(F). \tag{46}$$

The field factor is expressed as

$$\mathcal{W}_{\nu}(F) = \frac{\kappa}{2} \left( \frac{4\kappa^2}{F} \right)^{2n^* - 2n_{\xi} - |m| - 1} e^{-2\kappa^3/3F}. \tag{47}$$

The structure factor $G_{\nu}(\theta, \chi)$ is determined by an integral related to the ionizing orbital and a reference function, which has significant contribution only in the vicinity of the nuclei and is insensitive to the asymptotic behavior of the wavefunction:

$$G_{\nu}(\theta, \chi) = e^{-\kappa \mu_F} \int d\boldsymbol{r} \, \Omega_{\nu}^*(\hat{\boldsymbol{R}}^{-1}\boldsymbol{r}) \hat{V}_c \psi_0(\boldsymbol{r}), \tag{48}$$

evaluated within the mean-field framework (MF), where $\psi_0(\boldsymbol{r})$ represents the wavefunction of the ionizing orbital;

$$\boldsymbol{\mu} = -\int d\boldsymbol{r} \, \psi_0^*(\boldsymbol{r}) \boldsymbol{r} \psi_0(\boldsymbol{r}) \tag{49}$$

denotes the orbital dipole moment within the MF, with $\mu_F$ being its component along the direction of the external field;

$$\Omega_{\nu}(\boldsymbol{r}) = \sum_{l=|m|}^{\infty} \Omega_{lm}^{\nu}(\boldsymbol{r}) = \sum_{l=|m|}^{\infty} R_l^{\nu}(r) Y_{lm}(\hat{\boldsymbol{r}}) \tag{50}$$

constitutes a reference function that can be expanded into spherical harmonics, with its radial part given by

$$R_l^{\nu}(r) = \omega_l^{\nu} (\kappa r)^l \, e^{-\kappa r} \, M(l + 1 - n^*, 2l + 2, 2\kappa r), \tag{51}$$

where $M(a, b, x)$ represents the confluent hypergeometric function and

$$
\begin{aligned}
\omega_l^{\nu} = \ & (-1)^{l + (|m| - m)/2 + 1} \, 2^{l + 3/2} \, \kappa^{n^* - (|m| + 1)/2 - n_{\xi}} \\
& \times \sqrt{(2l + 1)(l + m)!(l - m)!(|m| + n_{\xi})! n_{\xi}!} \, \frac{l!}{(2l + 1)!} \\
& \times \sum_{k=0}^{\min(n_{\xi}, l - |m|)} \frac{\Gamma(l + 1 - n^* + n_{\xi} - k)}{k!(l - k)!(|m| + k)!(l - |m| - k)!(n_{\xi} - k)!}
\end{aligned}
\tag{52}
$$

serves as the normalization coefficient; $\hat{V}_c = \hat{V} + Z/r$ denotes the core potential with the Coulomb tail removed, where $Z$ signifies the asymptotic charge of the residual ion.

The effective potential $\hat{V}$ characterizes the interaction between the ionizing electron and the residual parent ion. We note that the hat notation is used to indicate that the potential operator is non-diagonal in the coordinate space. Under the Hartree-Fock formalism, the effective potential comprises three components, specifically the nuclear Coulomb potential ($V_{\text{nuc}}$), the direct ($V_{\text{d}}$) and exchange ($V_{\text{ex}}$) contributions of inter-electron interactions:

$$\hat{V} = V_{\text{nuc}} + V_{\text{d}} + \hat{V}_{\text{ex}}, \tag{53}$$

with

$$\begin{aligned}
V_{\text{nuc}}(\boldsymbol{r}) &= -\sum_{A=1}^{N_{\text{atm}}} \frac{Z_A}{|\boldsymbol{r} - \boldsymbol{R}_A|}, \\
V_{\text{d}}(\boldsymbol{r}) &= \sum_{i=1}^{N} \int \frac{\psi_i^*(\boldsymbol{r}')\psi_i(\boldsymbol{r}')}{|\boldsymbol{r} - \boldsymbol{r}'|} \mathrm{d}\boldsymbol{r}', \\
\hat{V}_{\text{ex}}\psi_0(\boldsymbol{r}) &= -\sum_{i=1}^{N} \psi_i(\boldsymbol{r}) \int \frac{\psi_i^*(\boldsymbol{r}')\psi_0(\boldsymbol{r}')}{|\boldsymbol{r} - \boldsymbol{r}'|} \langle \sigma_i | \sigma_0 \rangle \mathrm{d}\boldsymbol{r}',
\end{aligned} \tag{54}$$

where $N$ and $N_{\text{atm}}$ denote the number of electrons and atoms, respectively; $\psi_i(\boldsymbol{r})$ and $\sigma_i$ represent the molecular orbital and the spin state of the electron indexed by $i$, with $\langle \sigma_i | \sigma_j \rangle = 1$ indicating electrons $i$ and $j$ share the same spin state, and $\langle \sigma_i | \sigma_j \rangle = 0$ otherwise; $Z_A$ and $\boldsymbol{R}_A$ correspond to the nuclear charge and position of atom indexed by $A$.

Representing the rotated reference function in Eq. (48) through a linear combination of spherical harmonics using the Wigner-$D$ matrix facilitates efficient numerical evaluation of the structure factor using precomputed coefficients:

$$G_\nu(\theta, \chi) = \mathrm{e}^{-\kappa \mu_F} \sum_{l=|m|}^{\infty} \sum_{m'=-l}^{l} I_{lm'}^\nu d_{mm'}^l(\theta) \mathrm{e}^{-im'\chi}, \tag{55}$$

where the phase factor $\mathrm{e}^{-im\phi}$ in the expansion of $D_{mm'}^l(\phi, \theta, \chi) = \mathrm{e}^{-im\phi} d_{mm'}^l(\theta) \mathrm{e}^{-im'\chi}$ is omitted since it does not contribute to the final result, and the coefficient $I_{lm'}^\nu$ is defined as follows:

$$I_{lm'}^\nu = \int \mathrm{d}\boldsymbol{r}\, \Omega_{lm'}^{\nu_\kappa}(\boldsymbol{r}) \hat{V}_{\text{c}} \psi_0(\boldsymbol{r}). \tag{56}$$

The original WFAT provides the instantaneous tunneling ionization rate $w = \mathrm{d}W/\mathrm{d}t$, yet lacks the dependence on $k_{\text{t}}$. To adapt WFAT for preparing initial conditions of the electron samples, it is necessary to reformulate the original WFAT to incorporate a $k_{\text{t}}$-dependent rate. Here we adopt the $k_{\text{t}}$-dependence from MO-ADK [Eq. (A8)], which yields

$$\mathrm{d}W/\mathrm{d}t\mathrm{d}\boldsymbol{k}_{\text{t}} \propto k_{\text{t}}^{2|m|} \mathrm{e}^{-\kappa k_{\text{t}}^2/F} \tag{57}$$

under the small-$k_{\text{t}}$ limit. We modify the field factor $\mathcal{W}_\nu(F)$ according to the aforementioned $k_{\text{t}}$-dependence, resulting in the modified field factor

$$\begin{aligned}
\mathcal{W}_\nu(F, k_{\text{t}}) &= \mathcal{W}_\nu(F) \frac{(\kappa/F)^{|m|+1}}{|m|!} k_{\text{t}}^{2|m|} \mathrm{e}^{-\kappa k_{\text{t}}^2/F} \\
&\approx \frac{1}{2} \frac{\kappa^{|m|+2}}{F^{|m|+1}|m|!} \left( \frac{4\kappa^2}{F} \right)^{2n^*-2n_\xi-|m|-1} k_{\text{t}}^{2|m|} \exp\left[ -\frac{2}{3} \frac{(k_{\text{t}}^2 + \kappa^2)^{3/2}}{F} \right],
\end{aligned} \tag{58}$$

where the normalization coefficient is chosen such that

$$\mathcal{W}_\nu(F) = \int_0^\infty \mathcal{W}_\nu(F, k_{\text{t}}) 2\pi k_{\text{t}} \mathrm{d}k_{\text{t}}. \tag{59}$$

Through this approach, we derive the $k_{\text{t}}$-dependent rate as provided by the WFAT:

$$\frac{\mathrm{d}W^{\text{WFAT}}}{\mathrm{d}t\mathrm{d}\boldsymbol{k}_{\text{t}}} = \sum_\nu |G_\nu(\theta, \chi)|^2 \mathcal{W}_\nu[F(t), k_{\text{t}}]. \tag{60}$$

## B. Trajectory Evolution and Quantum Phase

Given the initial conditions, the electrons released from the tunnel exit subsequently evolve classically in the combination of Coulomb and laser fields, following a classical trajectory, and the scheme is named the *Classical Trajectory Monte Carlo (CTMC)*. In addition to the position and momentum, a quantum phase can be attributed to the evolving trajectories within frameworks such as the *Quantum Trajectory Monte Carlo (QTMC)* and the *Semiclassical Two-Step (SCTS) Model*. These models significantly preserve quantum effects in the final PMD, in contrast to the purely classical CTMC.

In this section we review the scheme of trajectory evolution and introduce the quantum phase methods available in `eTraj`.

### 1. Classical Trajectory Monte-Carlo (CTMC)

Within the CTMC framework, each electron, carrying an associated probability $W$, evolves along a classical trajectory until reaching a final momentum $\boldsymbol{p}_\infty = \boldsymbol{p}|_{t=\infty}$, which constitutes the primary observable of interest.

The electrons, emerging from the tunnel with distinct ionization times, initial positions, and momenta, evolve according to Hamilton's equations:

$$\dot{\boldsymbol{r}} = \boldsymbol{\nabla}_{\boldsymbol{p}} H, \quad \dot{\boldsymbol{p}} = -\boldsymbol{\nabla}_{\boldsymbol{r}} H, \tag{61}$$

and we use the Hamiltonian under the LG.

After the termination of the laser pulse, the electron interacts solely with the residual parent ion. At a distance from the parent ion, the electron experiences the Coulomb tail of the potential, and its Runge-Lenz vector

$$\boldsymbol{a} = \boldsymbol{p} \times \boldsymbol{L} - Z\boldsymbol{r}/r \tag{62}$$

can be considered asymptotically conserved. By leveraging the conservation of $\boldsymbol{a}$ alongside the angular momentum and energy, we derive the expression for the final momentum [84]:

$$\begin{aligned}
\boldsymbol{p}_\infty &= p_\infty \frac{p_\infty(\boldsymbol{L} \times \boldsymbol{a}) - \boldsymbol{a}}{1 + p_\infty^2 L^2}, \\
p_\infty^2/2 &= p^2/2 - Z/r, \\
\boldsymbol{L} &= \boldsymbol{r} \times \boldsymbol{p}, \\
\boldsymbol{a} &= \boldsymbol{p} \times \boldsymbol{L} - Z\boldsymbol{r}/r,
\end{aligned} \tag{63}$$

where $\boldsymbol{r}$ and $\boldsymbol{p}$ represent the electron's coordinate and momentum at any time subsequent to the termination of the laser pulse. This scheme applies to electrons with positive energy, which are capable of ultimately escaping the parent ion and being detected. Electrons with negative energy are presumed to be captured into Rydberg orbitals.

Electrons exhibiting comparable final momenta, i.e., those residing within the same bin of the final momentum grid, are aggregated by summing their respective probabilities: $W_{\boldsymbol{p}} = \sum_i W_i$. This aggregation results in the final momentum spectrum denoted by $W_{\boldsymbol{p}}$.

### 2. Quantum Trajectory Monte-Carlo (QTMC)

In contrast to the CTMC, the QTMC incorporates a quantum phase into each electron trajectory by employing the Feynman path-integral formalism [42]. This quantum phase is equivalent to the action $S_{\mathrm{traj}}$ defined in Eq. (19), with due consideration of the Coulomb potential.

The phase accumulates during the electron's excursion and is expressed as

$$\Phi^{\mathrm{QTMC}} = S_{\mathrm{traj}} = -\int_{t_{\mathrm{r}}}^{\infty} \left[ \frac{k^2}{2} + V(\boldsymbol{r}) + I_{\mathrm{p}} \right] \mathrm{d}t, \tag{64}$$

where $t_r$ signifies the instant when the electron exits the tunnel, and $\boldsymbol{k} = \dot{\boldsymbol{r}}$ represents the momentum. Ultimately, the PMD is determined by coherently summing the probability amplitudes corresponding to identical final momenta and taking the squared modulus of the resultant sum:

$$W_{\boldsymbol{p}} = \left| \sum_i \sqrt{W}_i \mathrm{e}^{\mathrm{i}\bar{S}_i} \right|^2, \tag{65}$$

where $\bar{S}_i$ encompasses the initial phase of the prefactor, the phase accumulated throughout the tunneling process, as well as the trajectory motion:

$$\bar{S} = \arg \mathcal{P}_{\boldsymbol{p}} + \Re S_{\mathrm{tun}} + S_{\mathrm{traj}}. \tag{66}$$

The phase $\Re S_{\mathrm{tun}}$ can be evaluated numerically but may be simplified if we adhere to the non-adiabatic-expansion scheme outlined in Sec. II A 2:

$$
\begin{aligned}
\Re S_{\mathrm{tun}} &\approx \Re \int_{t_s}^{t_r} \mathrm{d}\tau \left\{ \frac{1}{2} \left[ \boldsymbol{p} + \boldsymbol{A}(t_r) - \mathrm{i} t_i \boldsymbol{F}(t_r) + \frac{1}{2} t_i^2 \boldsymbol{F}'(t_r) \right]^2 + I_p \right\} \\
&= -[\boldsymbol{k}(t_r) \cdot \boldsymbol{F}(t_r)] \frac{t_i^2}{2} + o(t_i^2) \\
&\approx -\boldsymbol{k}_0 \cdot \boldsymbol{r}_0.
\end{aligned}
\tag{67}
$$

The last line of Eq. (67), i.e., $-\boldsymbol{k}_0 \cdot \boldsymbol{r}_0$, vanishes for the SFA-SPANE and ADK initial condition methods due to the absence of longitudinal initial momentum ($k_\parallel = 0$).

It is also noteworthy that in practical implementation, the upper limit of the integral in Eq. (64) need not extend to infinity. Since electrons arriving at the same final momentum share the same energy after the laser vanishes (at $t_f$), the integral

$$\int_{t_f}^{\infty} \left[ \frac{k^2}{2} + V(\boldsymbol{r}) + I_p \right] \mathrm{d}t \tag{68}$$

is uniform for all electrons with the same final momentum. Consequently, in numerical implementations, the upper limit of the phase integral in Eq. (64) can be set as the end of the laser pulse, i.e., $t_f$, leading to

$$S_{\mathrm{traj}}^{\mathrm{QTMC}} = -\int_{t_r}^{t_f} \left[ \frac{k^2}{2} + V(\boldsymbol{r}) + I_p \right] \mathrm{d}t. \tag{69}$$

### 3. Semiclassical Two-Steps (SCTS) Model

The SCTS model [45] enhances the quantum phase in the QTMC scheme, yielding

$$\Phi^{\mathrm{SCTS}} = \Re S_{\mathrm{tun}} + S_{\mathrm{traj}} = \underbrace{-\boldsymbol{k}_0 \cdot \boldsymbol{r}_0}_{\Re S_{\mathrm{tun}}} \underbrace{- \int_{t_0}^{\infty} \left[ \frac{k^2}{2} + V(\boldsymbol{r}) - \boldsymbol{r} \cdot \boldsymbol{\nabla} V(\boldsymbol{r}) + I_p \right] \mathrm{d}t}_{S_{\mathrm{traj}}}. \tag{70}$$

Compared to the QTMC phase given by Eq. (64), the SCTS phase from Eq. (70) differs in two key aspects: Firstly, there is the initial phase $-\boldsymbol{k}_0 \cdot \boldsymbol{r}_0$, which results from the tunneling process and is non-zero for non-adiabatic tunneling where $k_\parallel \neq 0$. Secondly, the integrand includes the $\boldsymbol{r} \cdot \boldsymbol{\nabla} V(\boldsymbol{r})$ term, which is absent in the QTMC formulation. These differences stem from the distinct theoretical foundations of the two methods: the QTMC phase is derived within the framework of first-order perturbation theory, whereas the SCTS formulation extends beyond this approximation. It should be noted that only the trajectory phase component of the SCTS model, represented by $S_{\mathrm{traj}}$ in Eq. (70), is adopted here. The tunneling phase $\Re S_{\mathrm{tun}}$ is intended to be incorporated during the preparation of initial conditions.

For the SCTS model, the phase integral in Eq. (70) over the interval $[t_f, \infty)$ cannot be simply neglected due to the presence of the $\boldsymbol{r} \cdot \boldsymbol{\nabla} V(\boldsymbol{r})$ term. However, the integral of this term can be reduced to an analytical expression in the case of a Coulomb potential, referred to as the post-pulse Coulomb phase:

$$
\begin{aligned}
S_{\mathrm{traj,f}}^{\mathrm{C}}(t_f) &= \int_{t_f}^{\infty} \boldsymbol{r} \cdot \boldsymbol{\nabla} V(\boldsymbol{r}) \mathrm{d}t \\
&= Z \int_{t_f}^{\infty} \frac{\mathrm{d}t}{r} \\
&= -\frac{Z}{p_{\infty}} \left[ \ln g + \sinh^{-1}\left( \frac{p_{\infty}}{g} \boldsymbol{r}_f \cdot \boldsymbol{p}_f \right) \right],
\end{aligned}
\tag{71}
$$

where $\boldsymbol{r}_f = \boldsymbol{r}(t_f)$, $\boldsymbol{p}_f = \boldsymbol{p}(t_f)$, and $g = \sqrt{1 + p_{\infty}^2 L^2} = \sqrt{1 + p_{\infty}^2 (\boldsymbol{r}_f \times \boldsymbol{p}_f)^2}$.

In this manner, we derive the expression for the SCTS trajectory phase suitable for numerical implementation:

$$
S_{\mathrm{traj}}^{\mathrm{SCTS}} = I_{\mathrm{p}} t_{\mathrm{r}} - \int_{t_{\mathrm{r}}}^{t_f} \left[ \frac{k^2}{2} + V(\boldsymbol{r}) - \boldsymbol{r} \cdot \boldsymbol{\nabla} V(\boldsymbol{r}) \right] \mathrm{d}t + S_{\mathrm{traj,f}}^{\mathrm{C}}(t_f).
\tag{72}
$$

## III. PROGRAM ARCHITECTURE AND MANUAL

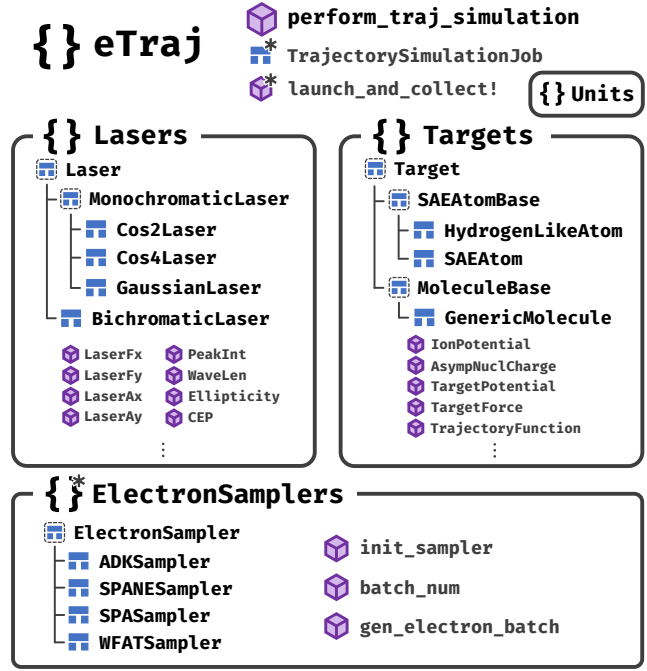

FIG. 1. Architectural structure diagram of `eTraj`.

The architectural structure of `eTraj` is illustrated in Fig. 1. The program consists of four primary components: `Targets`, `Lasers`, `ElectronSamplers`, and `TrajectorySimulationJob`. Each component serves a critical function within the library, forming an essential link of the overall workflow.

### A. `Lasers` Module

A conventional monochromatic laser field comprises a carrier wave $\cos(\omega t + \phi)$ modulated by an envelope function $f_{\mathrm{env}}(t)$ (normalized to unity at its peak). Given the amplitude of the vector potential $A_0$, the time-dependent vector potential of the laser is expressed as follows. We assume the laser propagates along the $z$-direction, with the $x$-axis as the principal axis of polarization:

$$\boldsymbol{A}(t) = A_0 f_{\mathrm{env}}(t)[\cos(\omega t + \phi)\hat{\boldsymbol{x}} + \varepsilon \sin(\omega t + \phi)\hat{\boldsymbol{y}}], \tag{73}$$

where $\omega$ represents the laser angular frequency, $\phi$ denotes the carrier-envelope phase (CEP), and $\varepsilon$ is the ellipticity.

The `Lasers` module provides several pre-implemented monochromatic laser objects, including `Cos4Laser`, `Cos2Laser`, and `GaussianLaser`, which are distinguished by their respective envelope functions $f_{\mathrm{env}}(t)$. All of these laser types are derived from the base type `MonochromaticLaser`.

The `Cos4Laser`'s vector potential has a $\cos^4$-shaped envelope function:

$$f_{\mathrm{env}}^{\cos 4} = \begin{cases} \cos^4\left[\dfrac{\omega(t - t_0)}{2N}\right], & -NT/2 \le t - t_0 \le NT/2, \\ 0, & \text{otherwise,} \end{cases} \tag{74}$$

where $N$ is the total cycle number, $T = 2\pi/\omega$ is the period and $t_0$ the peak time.

The `Cos2Laser` has a $\cos^2$-shaped envelope function, similar to that of the `Cos4Laser`:

$$f_{\text{env}}^{\cos2} = \begin{cases} \cos^2\left[\dfrac{\omega(t-t_0)}{2N}\right], & -NT/2 \le t - t_0 \le NT/2, \\ 0, & \text{otherwise.} \end{cases} \tag{75}$$

The `GaussianLaser` has a Gaussian-shaped envelope function, which is the most commonly used:

$$f_{\text{env}}^{\text{Gauss.}} = e^{-(t-t_0)^2/2\tau_\sigma^2} = e^{-\ln 2 \cdot (t-t_0)^2/\tau_{\text{FWHM}}^2}, \tag{76}$$

where $\tau_\sigma$ is the temporal width of the laser and $\tau_{\text{FWHM}} = 2\sqrt{\ln 2}\,\tau_\sigma$ denotes the laser's intensity profile's temporal FWHM (full-width at half maxima).

The constructor methods for monochromatic laser objects are defined with the following signatures:

```
Cos4Laser(peak_int, wave_len|ang_freq, cyc_num|duration, ellip [,azi=0] [,cep=0] [,t_shift=0])
Cos2Laser(peak_int, wave_len|ang_freq, cyc_num|duration, ellip [,azi=0] [,cep=0] [,t_shift=0])
GaussianLaser(peak_int, wave_len|ang_freq, spread_cyc_num|spread_duration|FWHM_duration, ellip [,azi=0] [,cep=0]
↪ [,t_shift=0])
```

where parameters enclosed in square brackets (e.g., `[azi=0]`) denote optional arguments, where default values are applied if the parameters are not explicitly provided; the vertical bar (|) signifies mutually exclusive alternatives for a given parameter (e.g., `wave_len|ang_freq` indicates that either the wavelength or the angular frequency may be specified, but not both simultaneously). For a detailed description of the parameters, see Table I.

| Parameter | Expression | Description | Unit |
|---|---|---|---|
| `peak_int` | $I_0 = A_0^2\omega^2(1+\varepsilon^2)$ | Peak intensity | W/cm$^2$ |
| `wave_len` | $\lambda = c_0 T$ | Wavelength | nm |
| `ang_freq` | $\omega$ | Angular frequency | a.u. |
| `ellip` | $\varepsilon$ | Ellipticity ($-1 \le \varepsilon \le 1$, 0 indicates linear polarization and $\pm 1$ indicates circular polarization) | — |
| `cyc_num` | $N$ | [`Cos2Laser` and `Cos4Laser`] Number of total cycles | — |
| `duration` | $NT$ | [`Cos2Laser` and `Cos4Laser`] Duration | a.u. |
| `spread_cyc_num` | $N_\sigma = \tau_\sigma/T$ | [`GaussianLaser`] Temporal width converted to cycle numbers | — |
| `spread_duration` | $\tau_\sigma$ | [`GaussianLaser`] Temporal width | a.u. |
| `FWHM_duration` | $\tau_{\text{FWHM}}$ | [`GaussianLaser`] Temporal FWHM of the intensity profile | a.u. |
| `azi` | — | Azimuth angle of the polarization's principle axis relative to the $x$ axis | rad |
| `cep` | $\phi$ | Carrier-Envelope-Phase | rad |
| `t_shift` | $t_0$ | Time shift relative to the peak | a.u. |

TABLE I. Input parameters of the constructor methods of the monochromatic laser objects in the `Lasers` module.

Apart from monochromatic lasers, the `BichromaticLaser` which combines two `MonochromaticLasers` is also implemented. A `BichromaticLaser` is initialized by the method:

```
BichromaticLaser(l1::MonochromaticLaser, l2::MonochromaticLaser [,delay=0])
```

with `delay` being the time delay of `l2` relative to `l1`.

The program imports the `Unitful.jl` package, which facilitates seamless unit conversion. Users can input an `Unitful.Quantity` with arbitrary units, as the program automatically performs the necessary conversions. Additionally, commonly used units are exported in the submodule `eTraj.Units` and are readily accessible upon importing the module, see note [85].

The example of usage is shown below, which runs in the Julia REPL (run-eval-print-loop) terminal [86]:

```julia
julia> using eTraj.Lasers

julia> l = Cos2Laser(peak_int=4e14, wave_len=800.0, cyc_num=2.0, ellip=0.0)      # without units, following default units
[MonochromaticLaser] Envelope cos², peak intensity 4.0e+14 W/cm², wavelen=800 nm, 2 cycle(s), ε=0 [linearly polarized]

julia> using eTraj.Units    # import the Units submodule to access units

julia> l = Cos4Laser(peak_int=0.4PW/cm^2, ang_freq=1.5498eV, duration=26.7fs, ellip=1.0, cep=90°)
[MonochromaticLaser] Envelope cos⁴, peak intensity 4.0e+14 W/cm², wavelen=800.00 nm, 10.01 cycle(s), ε=1 [circularly polarized], CEP=0.50 π

julia> l = GaussianLaser(peak_int=4e14W/cm^2, wave_len=400nm, FWHM_duration=20fs, ellip=-1)
[MonochromaticLaser] Envelope Gaussian, peak intensity 4.0e+14 W/cm², wavelen=400 nm, temporal width 9.00 cycle(s) [FWHM 20.00 fs], ε=-1
↪ [circularly polarized]

julia> l = BichromaticLaser(l1=Cos4Laser(peak_int=1.0PW/cm^2, wave_len=800nm, cyc_num=10, ellip=1), l2=Cos4Laser(peak_int=1.0PW/cm^2,
↪ wave_len=400nm, cyc_num=20, ellip=-1), delay=0.5fs)
[BichromaticLaser] delay Δt = 20.67 a.u. (0.50 fs)
├ [MonochromaticLaser] Envelope cos⁴, peak intensity 1.0e+15 W/cm², wavelen=800 nm, 10 cycle(s), ε=1 [circularly polarized]
└ [MonochromaticLaser] Envelope cos⁴, peak intensity 1.0e+15 W/cm², wavelen=400 nm, 20 cycle(s), ε=-1 [circularly polarized]
```

We note that the constructor methods should be invoked with keyword arguments, as is shown in the example.

### B. `Targets` Module

The `Targets` module implements the abstraction of targets and provides parameters of some commonly-used targets.

The `HydrogenLikeAtom` and `SAEAtom` are both subtypes of the `SAEAtomBase` base type, which represents an atom under the SAE approximation. The distinguishing characteristic between these types lies in their potential functions describing the residual ion after electron ionization.

The `HydrogenLikeAtom`'s potential function is of the form:

$$V(r) = -\frac{Z}{\sqrt{r^2 + a}}, \tag{77}$$

with $a$ the soft-core parameter which is tuned to match the atom's ionization potential, and is also adopted to avoid singularity of the potential during numerical simulation.

The `SAEAtom`'s potential function is adopted from Tong's model [87]:

$$V(r) = -\frac{Z + a_1 \mathrm{e}^{-b_1 r} + a_2 r \mathrm{e}^{-b_2 r} + a_3 \mathrm{e}^{-b_3 r}}{\sqrt{r^2 + a}}, \tag{78}$$

where $a_i$ and $b_i$ are tunable parameters to fit the effective potential, which can be obtained using the procedure detailed in Ref. [87].

The signatures of the constructor methods of `HydrogenLikeAtom` and `SAEAtom` are listed below:

```
    HydrogenLikeAtom(Ip, Z [,l=0] [,m=0] [,asymp_coeff=:hartree|<coeff>] [,quan_ax_θ=0] [,quan_ax_ϕ=0] [,soft_core=1e-10]
↪ [,name])
    SAEAtom(Ip, Z [,l=0] [,m=0] [,asymp_coeff=:hartree|<coeff>] [,quan_ax_θ=0] [,quan_ax_ϕ=0] [,a1,b1,a2,b2,a3,b3]
↪ [,soft_core=1e-10] [,name])
```

A detailed description of the parameters is listed in Table II. We note that ionization starting from excited states is possible by adjusting the relevant parameters $I_{\rm p}, l, m$ ... during initialization.

The definitions of `HydrogenLikeAtom` and `SAEAtom` do not prohibit inconsistent combinations of ionization potential ($I_{\rm p}$) and potential parameters, which may appear contradictory. This arises because these properties are related to two processes that are relatively independent within the trajectory simulation framework: $I_{\rm p}$ is solely involved in setting the initial conditions of the photoelectrons, whereas the potential $V(\boldsymbol{r})$ affects the subsequent trajectory-evolution dynamics. Therefore, for custom-defined atoms, it is recommended to verify the consistency between $I_{\rm p}$ and the potential—ensuring that $I_{\rm p}$ corresponds to the ionization energy of the desired eigenstate of the potential—to achieve better alignment with the TDSE results.

To facilitate user convenience, a set of predefined configurations for commonly used atoms is provided, accessible via the `get_atom` method. The available keys can be retrieved by calling the `get_available_atoms()` method.

Finally, we present an example of using `HydrogenLikeAtom` and `SAEAtom` in REPL:

```julia
julia> using eTraj.Targets

julia> t = HydrogenLikeAtom(Ip=0.5, Z=1, name="H")
[HydrogenLikeAtom] Atom H, Ip=0.5000 (13.61 eV), Z=1

julia> using eTraj.Units

julia> t = SAEAtom(Ip=12.13eV, Z=1, l=1, a1=51.35554, b1=2.111554, a2=-99.92747, b2=3.737221, a3=1.644457, b3=0.4306465,
↪ asymp_coeff=1.3, name="Xe")
[SAEAtom] Atom Xe (p orbital, m=0), Ip=0.4458 (12.13 eV), Z=1

# example of `get_atom`

julia> t = get_atom("He1p")
[HydrogenLikeAtom] Atom He⁺, Ip=1.0000 (27.21 eV), Z=2

julia> t = get_atom("Xe"; m=1, quan_ax_θ=90°, quan_ax_ϕ=0°) # other parameters can be passed via keyword arguments
[SAEAtom] Atom Xe (p orbital, m=1), Ip=0.4458 (12.13 eV), Z=1, θϕ=(90.0°,0.0°)
```

| Parameter | Description |
|---|---|
| `Ip` | Ionization potential (default unit is a.u.) |
| `Z` | Asymptotic charge of the residual ion |
| `l` | Angular quantum number |
| `m` | Magnetic quantum number |
| `asymp_coeff` | Asymptotic coefficient $C_{\kappa l}$, setting `:hartree` indicates automatic calculation using Eq. (14) |
| `quan_ax_θ` | Quantization axis' polar angle in the LF |
| `quan_ax_ϕ` | Quantization axis' azimuth angle in the LF |
| `name` | Target's name |
| `soft_core` | Soft-core parameter |
| `a1,b1,a2,b2,a3,b3` | [`SAEAtom`] Tunable parameter to fit the effective potential |

TABLE II. Input parameters of the constructor method of `HydrogenLikeAtom` and `SAEAtom` in the `Targets` module.

For molecular targets, the implementation includes the `GenericMolecule` type, which incorporates a more sophisticated structure. A `GenericMolecule` stores information about the atoms that form the molecule, together with their coordinates, as well as the asymptotic coefficients [$C_{lm}$ in Eq. (38)] and WFAT's integral coefficients [Eq. (56)], which are obtained using other quantum chemistry packages.

There are two ways to initialize a `GenericMolecule`: build from zero or from an existing file, see the definition below. The description of the corresponding parameters is listed in Table III.

```
GenericMolecule(atoms, atom_coords [,charge=0] [,spin=0] [,name] [,rot_α=0] [,rot_β=0] [,rot_γ=0])
LoadMolecule(ext_data_path; [rot_α=0] [,rot_β=0] [,rot_γ=0])
```

| Parameter | Description |
|---|---|
| `atoms` | Atoms in the molecule, stored as a `Vector` of `String` |
| `atom_coords` | Atoms' coordinates in the molecule, stored as a $N \times 3$ matrix (default unit is Å) |
| `charge` | Total charge number (i.e., $Z - 1$) |
| `spin` | Total spin (each unpaired electron contributes 1/2) |
| `name` | Molecule's name |
| `rot_α`,`rot_β`,`rot_γ` | Euler angles ($z - y' - z''$ convention) specifying the molecule's orientation in LF (default unit is radian) |

TABLE III. Input parameters of the initialization methods of `GenericMolecule` in the `Targets` module.

For quantum chemistry calculations, we have implemented a computational scheme in `PySCFMolecularCalculator` utilizing PySCF [88], which is compatible with Linux and macOS platforms (Windows users can access this functionality through the Windows Subsystem for Linux (WSL)). The calculation scheme of the WFAT structure factor is adopted from `PyStructureFactor` [83]. Future extension is possible by implementing the supertype `MolecularCalculatorBase`. Since there are some predefined molecules available via `get_mol`, we are not going to detail on the manual of running the calculation in the text.

The example of initializing and calculating the essential data of `GenericMolecule` in REPL is presented as follows:

```
julia> using eTraj.Targets, eTraj.Units

julia> mol = GenericMolecule(atoms=["O","C","O"], atom_coords=[0 0 -1.1600; 0 0 0; 0 0 1.1600]*Å, charge=0, name="Carbon
↪ Dioxide (CO₂)")
[GenericMolecule] Carbon Dioxide (CO₂)

julia> MolInitCalculator!(mol, basis="cc-pVTZ")
[ Info: [PySCFMolecularCalculator] Running molecular calculation ...

julia> MolCalcAsympCoeff!(mol, 0); MolCalcAsympCoeff!(mol, -1)
[ Info: [PySCFMolecularCalculator] Running calculation of asymptotic coefficients ... (ionizing orbital HOMO)
[ Info: [PySCFMolecularCalculator] Running calculation of asymptotic coefficients ... (ionizing orbital HOMO-1)

julia> MolCalcWFATData!(mol, 0); MolCalcWFATData!(mol, -1)
[ Info: [PySCFMolecularCalculator] Running calculation of WFAT structure factor data ... (ionizing orbital HOMO)
[ Info: [PySCFMolecularCalculator] Running calculation of WFAT structure factor data ... (ionizing orbital HOMO-1)

julia> MolSaveDataAs!(mol, "Molecule_CO2.jld2")  # save the data to a file
[ Info: [GenericMolecule] Data saved for molecule Carbon Dioxide (CO₂) at `Molecule_CO2.jld2`.

julia> mol_ = LoadMolecule("Molecule_CO2.jld2")  # load from saved file
[GenericMolecule] Carbon Dioxide (CO₂)
```

```
Asymp coeff of HOMO-1 & HOMO available
WFAT data of HOMO-1 & HOMO available
#          E (Ha)  occp
⋮    ⋮       ⋮       ⋮⋮
13 LUMO+1   0.207   ——
12 LUMO     0.175   ——
11 HOMO    -0.542   -↑↓-
10 HOMO-1  -0.542   -↑↓-
9  HOMO-2  -0.714   -↑↓-
8  HOMO-3  -0.714   -↑↓-
⋮    ⋮       ⋮       ⋮⋮
```

## C. Electron Sampling and Trajectory Simulation

The `ElectronSamplers` module provides means of generating initial electron samples using different initial condition methods. The `ElectronSampler` is an abstract supertype, with `ADKSampler`, `SPANESampler`, `SPASampler` and `WFATSampler` being its subtypes. When the user starts a trajectory simulation job by invoking `eTraj.perform_traj_simulation`, the method would further call the `init_sampler` method, which would assign the corresponding type of `ElectronSampler` in the background. These operations execute internally, maintaining the `ElectronSamplers` module as a private component not exposed for direct public access.

The `eTraj.perform_traj_simulation` method serves as a public entrance to performing a trajectory simulation. The method automatically detects available computational threads (specified via command-line argument "`-t <thread_num>`" at Julia startup, e.g., "`julia-1.11 -t 4`") and executes trajectory simulations in parallel. Table IV details on the input parameters of the method.

The following describes the operational workflow of the `eTraj.perform_traj_simulation` method:

- Initially, the method creates an `eTraj.TrajectorySimulationJob` instance to store essential parameters, with the electron sampler assigned based on the specified `init_cond_method`.

- Subsequently, it performs repeated invocations of `eTraj.launch_and_collect!`, where each invocation processes a batch of electrons released at time $t_r$ with varying transversal momenta $\boldsymbol{k}_t$ using the appropriate simulation scheme.

  - The sampling methodology is determined by the `sample_monte_carlo` parameter: when set to `true`, electron batch times $t_r$ and momenta $\boldsymbol{k}_t$ are randomly sampled within intervals defined by `sample_t_intv`, `mc_kd_max`, and `mc_kz_max`; otherwise, initial conditions are sampled equidistantly with intervals specified by `sample_t_intv`, `ss_kd_max`, and `ss_kz_max`.

- Following the generation of initial electron batches, the `eTraj.launch_and_collect!` method conducts classical trajectory simulations (optionally including phase calculations) and aggregates final momenta on a predefined grid. These simulations are implemented using the `OrdinaryDiffEq.jl` package [89].

  - The momentum grid's dimensions and resolution are configured through the `final_p_max` and `final_p_num` parameters.

- Upon completion, the method generates an output file (in either JLD2 or HDF5 format) containing the PMD along with other essential simulation data.

The output file is a Julia JLD2 file, which is compatible with the HDF5 data format, and can be opened by the `JLD2.jl` package:

```julia
julia> using JLD2

julia> file = jldopen("ADK-CTMC_4e14_800nm_cos4_2cyc_CP.jld2")
JLDFile .../ADK-CTMC_4e14_800nm_cos4_2cyc_CP.jld2 (read-only)
    ├─ info
    ├─ params_text          # parameters stored in YAML format
    ├─ params               # parameters stored in a Julia `Dict`
    ├─ px                   # coordinates of the `momentum_spec` on x-axis
    ├─ py                   # coordinates of the `momentum_spec` on y-axis
    ├─ momentum_spec        # PMD data stored in a Julia `Array`
    ├─ ion_prob             # total ionization probability
    ├─ ion_prob_uncollected # ionization probability of discarded electrons
    └─ num_effective_traj   # total number of effective trajectories

julia> print(file["params_text"])
init_cond_method: ADK
laser:
    type: Cos4Laser
    peak_int: 4.0e14
    wave_len: 800.0
    ⋮
target:
    type: HydrogenLikeAtom
    Ip: 0.5
    nucl_charge: 1
    ⋮
sample_t_intv: (-100, 100)
sample_t_num: 20000
sample_monte_carlo: false
    ⋮
```

| Parameter | Description | Default |
|---|---|---|
| *Required parameters* | | |
| `init_cond_method` | Method used to determine the initial conditions of electrons. Candidates: `:ADK`, `:SPA` (SFA-SPA), `:SPANE` (SFA-SPANE) for targets of type `SAEAtomBase` or `MoleculeBase`; `:WFAT` for `MoleculeBase` targets. | |
| `laser` | A `Lasers.Laser` object which stores parameters of the laser field. | |
| `target` | A `Targets.Target` object which stores parameters of the target. | |
| `dimension` | Dimensionality of the simulation (`2` or `3`). 2D simulation is carried out in the $x-y$ plane. | |
| `sample_t_intv` | Time interval for sampling initial electrons. Format: `(start,stop)`. Default unit: a.u. | |
| `sample_t_num` | Number of time samples. | |
| `traj_t_final` | Final time of each trajectory simulation. Default unit: a.u. | |
| `final_p_max` | Boundaries of final momentum grid. Grid ranges from `-pxMax` to `+pxMax` in the $x$ direction, and the same for $y$ and $z$ directions. Format: `(pxMax,pyMax[,pzMax])` | |
| `final_p_num` | Numbers of final momentum grid points. If a value is `1`, electrons will be collected regardless of the momentum on that dimension. Format: `(pxNum,pyNum[,pzNum])` | |
| *Required parameters for step sampling (SS) methods* (`sample_monte_carlo=false`) | | |
| `ss_kd_max` | Boundary of $k_\perp$ samples (in a.u.). $k_\perp$ ranges from `-ss_kd_max` to `+ss_kd_max`. | |
| `ss_kd_num` | Number of $k_\perp$ samples. | |
| `ss_kz_max` | [3D only] Boundary of $k_z$ samples (in a.u.). $k_z$ ranges from `-ss_kz_max` to `+ss_kz_max`. | |
| `ss_kz_num` | [3D only] Number of $k_z$ samples. | |
| *Required parameters for Monte-Carlo (MC) sampling methods* (`sample_monte_carlo=true`) | | |
| `mc_kt_num` | Number of $k_t$ samples in a single time sample. | |
| `mc_kd_max` | Boundary of $k_\perp$ (in a.u.). $k_\perp$ ranges from `-mc_kd_max` to `+mc_kd_max`. | |
| `mc_kz_max` | [3D only] Boundary of $k_z$ (in a.u.). $k_z$ ranges from `-mc_kz_max` to `+mc_kz_max`. | |
| *Optional parameters* | | |
| `traj_phase_method` | Method used to determine classical trajectories' phase. Candidates: `:CTMC`, `:QTMC`, and `:SCTS`. Note: The `WFAT` initial condition only supports `:CTMC`. | `:CTMC` |
| `traj_rtol` | Relative error tolerance for solving classical trajectories. | `1e-6` |
| `output_fmt` | Output file format. Candidates: `:jld2` (JLD2) and `:h5` (HDF5). | `:jld2` |
| `output_compress` | Determines whether output files are compressed or not. Note: For JLD2 output format, compression requires explicit installation of the `CodecZlib.jl` package. | `:true` |
| `output_path` | Path to output file. | |
| `sample_cutoff_limit` | Probability cutoff limit for sampled electrons. Electrons with probabilities lower than the limit would be discarded. | `1e-16` |
| `sample_monte_carlo` | Determines whether Monte-Carlo sampling is used when generating electron samples. | `false` |
| *Optional parameters for atomic SFA-SPA, SFA-SPANE and ADK methods* | | |
| `rate_prefix` | Prefix of the exponential term in the ionization rate. `:Exp` indicates no prefix; `:Pre` and `:PreCC` indicates inclusion of the prefactor $\mathcal{P}$ with or without the Coulomb correction $C^{\mathrm{CC}}$; `:Jac` indicates inclusion of the Jacobian factor $J$ which is related to the sampling method; `:Full` is equivalent to `Set([:PreCC,:Jac])`. To combine `:Pre` and `:Jac`, pass `Set([:Pre,:Jac])`. | `:Full` |
| *Optional parameters for target* `MoleculeBase` | | |
| `mol_orbit_ridx` | Index of selected orbital relative to the HOMO (e.g., `0` indicates HOMO, and `-1` indicates HOMO-1.) For open-shell molecules, according to $\alpha/\beta$ spins, should be passed in format `(spin, idx)` where for $\alpha$ orbitals `spin=1` and for $\beta$ orbitals `spin=2`. | `0` |
| *Optional parameters for interface* | | |
| `show_progress` | Whether to display progress bar. | `true` |

TABLE IV. Input parameters of the `eTraj.perform_traj_simulation` method.

## IV. ILLUSTRATIVE EXAMPLES

This section presents selected examples demonstrating key applications of `eTraj`, illustrating its capabilities across representative scenarios. The example and the corresponding post-processing codes are provided in the `examples/` directory.

The implementation consists of two types of scripts: simulation scripts (named "`test_*.jl`") that execute trajectory simulations and generate JLD2 output files, and visualization scripts (named "`plot_*.jl`") that process these files to produce figures. To run the plotting scripts, the user must install and successfully build the `Plots.jl` and `PyPlot.jl` packages.

The execution time of the program is primarily determined by the number of *effective* trajectories, defined as those not filtered out due to probabilities below the `sample_cutoff_limit` threshold. Additional factors influencing computational efficiency include the choice of initial condition and phase methods, the system's dimensionality (specified by the `dimension` parameter), the relative error tolerance of the ODE integrator (specified by the `traj_rtol` parameter), and the occurrence of electron recollisions, which necessitate finer time steps for accurate resolution. A summary of the execution times for the presented examples is provided in Table V.

| Example | Configuration | Cores | Eff. traj. | Exec. time | Speed |
|---|---|---|---|---|---|
| test_2cycs_CP | ADK-CTMC | 12 | 54.2M | 3′53″ | 4.3 µs/traj |
| | ADK-CTMC | 8 | 54.2M | 4′43″ | 5.2 µs/traj |
| | ADK-CTMC | 6 | 54.2M | 5′32″ | 6.1 µs/traj |
| | ADK-CTMC | 4 | 54.2M | 7′23″ | 8.2 µs/traj |
| | ADK-CTMC | 2 | 54.2M | 11′22″ | 12.5 µs/traj |
| | ADK-CTMC | 1 | 54.2M | 20′49″ | 23.0 µs/traj |
| | SPANE-CTMC | 12 | 69.7M | 4′25″ | 3.8 µs/traj |
| | SPA-CTMC | 12 | 83.9M | 22′38″ | 16.2 µs/traj |
| test_8cycs_CP | ADK-CTMC | 12 | 131M | 15′20″ | 7.0 µs/traj |
| | ADK-QTMC | | | 17′59″ | 8.2 µs/traj |
| | ADK-SCTS | | | 20′32″ | 9.4 µs/traj |
| test_1cyc_CP | ADK-CTMC | 12 | 22.8M | 1′42″ | 4.5 µs/traj |
| | ADK-QTMC | | | 2′00″ | 5.3 µs/traj |
| | ADK-SCTS | | | 2′02″ | 5.4 µs/traj |
| test_Bichromatic_CCP | $I_0 = 1 \times 10^{14}$ W/cm$^2$ | 12 | 18.7M | 3′54″ | 12.5 µs/traj |
| | $I_0 = 3 \times 10^{14}$ W/cm$^2$ | | 29.0M | 6′04″ | 12.5 µs/traj |
| | $I_0 = 5 \times 10^{14}$ W/cm$^2$ | | 32.9M | 7′00″ | 12.8 µs/traj |
| | $I_0 = 7 \times 10^{14}$ W/cm$^2$ | | 35.2M | 7′35″ | 12.9 µs/traj |
| test_Molecules | $H_2$ HOMO | 12 | 14.7M | 1′45″ | 7.1 µs/traj |
| | CO HOMO | | 16.2M | 1′47″ | 6.6 µs/traj |
| | $O_2$ α-HOMO | | 15.7M | 1′45″ | 6.7 µs/traj |
| | $O_2$ α-HOMO-1 | | 15.3M | 1′42″ | 6.7 µs/traj |
| | $C_6H_6$ HOMO | | 22.3M | 2′21″ | 6.3 µs/traj |
| | $C_6H_6$ HOMO-1 | | 22.0M | 2′19″ | 6.3 µs/traj |

TABLE V. Execution time summary for examples in Section IV. The execution time does not include the package loading or precompilation time. The examples are run on an AMD Ryzen 9 7950X CPU with 12 available cores and 32 GB of RAM, on WSL Ubuntu 22.04 LTS, Julia version 1.11.1.

## A. Attoclock Experiment and the Influence of Initial Condition Methods

This example is adapted from Ref. [68].

The attoclock experiment employs an ultra-short circularly polarized pulse to investigate ultrafast attosecond dynamics, particularly tunneling time delay phenomena. Here, we simulate an attoclock experiment using three initial condition methods—ADK, SFA-SPANE, and SFA-SPA—to examine how non-adiabatic effects influence the attoclock signal. We choose CTMC as the phase method because the quantum interference effect is not significant in this example. In our theoretical framework, the simulation schemes are named after "ADK-CTMC", "(SFA-)SPANE-CTMC" and "(SFA-)SPA-CTMC", respectively.

```
                              ──── examples/test_2cycs_CP.jl ────
1   using eTraj
2   using eTraj.Targets, eTraj.Lasers, eTraj.Units
3
4   l = Cos4Laser(peak_int=0.4PW/cm^2, wave_len=800.0nm, cyc_num=2, ellip=1.0)
5   t = get_atom("H")
6
7   for init_cond in [:ADK, :SPANE, :SPA]
8       perform_traj_simulation(
9           init_cond_method    = init_cond,
10          traj_phase_method   = :CTMC,
11          laser               = l,
12          target              = t,
13          dimension           = 2,                # 2D simulation, x-y plane only
14          sample_t_intv       = (-100,100),       # equivalent to `(-2.42fs, 2.42fs)`
15          sample_t_num        = 20000,            # will sample 20000 equidistant time points between -100 and 100 a.u.
16          traj_t_final        = 120,              # the traj end at 120 a.u., equivalent to `2.90fs`
17          final_p_max         = (2.5,2.5),        # the momentum spec collection grid's border (-2.5 to +2.5 a.u.)
18          final_p_num         = (500,500),        # the momentum spec collection grid's size (500×500)
19          ss_kd_max           = 2.0,
20          ss_kd_num           = 10000,            # will sample 10000 equidistant k⊥ points between -2 to +2 a.u.
21          output_path         = "$(init_cond)-CTMC_4e14_800nm_cos4_2cyc_CP.jld2"
22      )
23  end
```

The PMD results are presented in Fig. 2. Owing to the exponential dependence of the ionization rate on the field strength, the PMD exhibits a crescent-shaped structure near the peak of the negative vector potential $-\boldsymbol{A}(t)$. In the adiabatic tunneling regime, corresponding to the ADK initial condition, the trajectory of $-\boldsymbol{A}(t)$ is expected to align with the median of the crescent-shaped structure. In contrast, for non-adiabatic tunneling, the distribution of the initial transverse momentum $\boldsymbol{k}_t$ at the tunnel exit is centered at a nonzero value. This leads to an expansion of the crescent-shaped structure and an enhancement of the overall ionization probability, as evident in the figure. Furthermore, the PMDs obtained using the SFA-SPANE and SFA-SPA initial conditions exhibit similar shapes and total ionization probabilities. This demonstrates the advantage of the SFA-SPANE approach in preserving non-adiabatic effects while significantly reducing computational costs compared to the SFA-SPA method.

## B. Interaction with Linearly-polarized Pulses and the Influence of Quantum Phase

This example is adapted from Ref. [45], which introduced the initial SCTS model.

When atoms interact with intense linearly-polarized pulses, they may absorb excess photons beyond the ionization threshold, resulting in characteristic ring-like structures in the PMD known as above-threshold ionization (ATI). While ATI was initially understood through the "multi-photon absorption" framework, it can also be interpreted within the tunneling regime as the result of intercycle interference between electron wave packets that tunnel through the potential barrier at successive peaks of the laser field. This allows for reproduction of the ATI structures using the semiclassical trajectory-based methods.

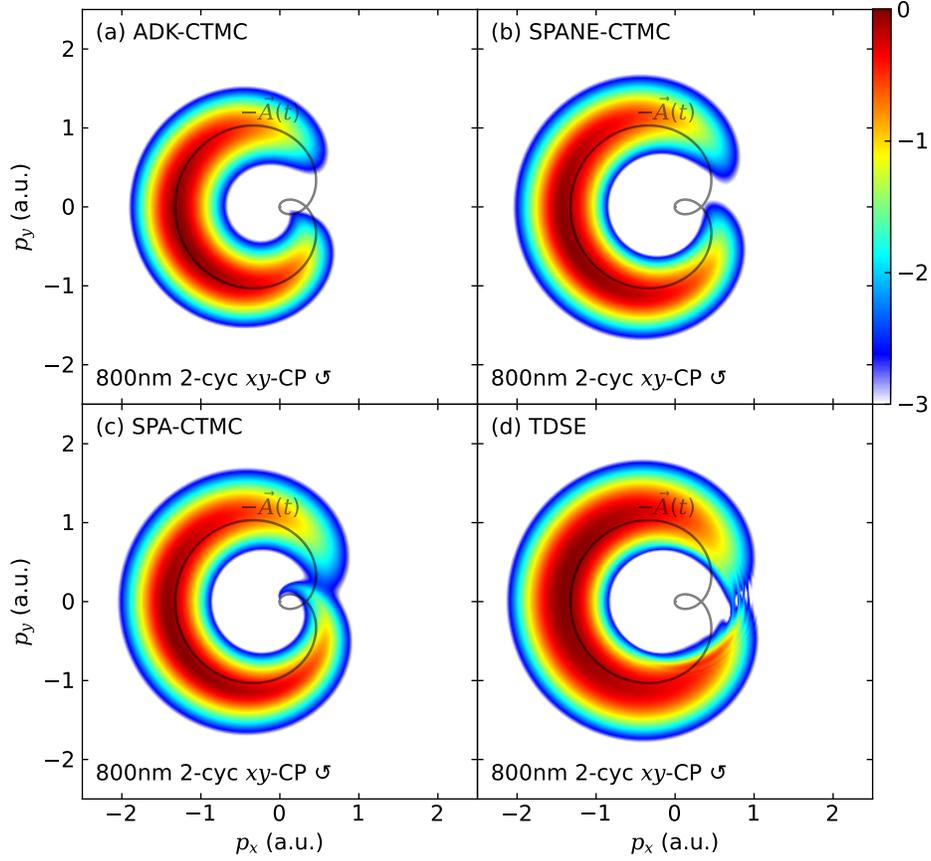

FIG. 2. Signals of an attoclock experiment obtained from CTMC simulation with (a) the ADK, (b) the SFA-SPANE, and (c) the SFA-SPA initial conditions, as well as (d) the TDSE simulation (logarithmic scale, each normalized to the maximum value). The laser is a 2-cycle circularly-polarized 800-nm pulse with a $\cos^4$ shape and peak intensity of 0.4 PW/cm$^2$, which polarizes in the $x - y$ plane. The target is a hydrogen atom initially at the ground state. In each sub-figure, the trace of the negative vector potential, $-\boldsymbol{A}(t)$, is depicted as a gray line for reference.

Beyond ATI rings, the PMD also exhibits intriguing low-energy features characterized by fan-like structures. This structure is contributed mainly by electrons with an angular momentum close to a specific value $L_0$ and are hence predictable theoretically [90, 91].

The following two code snippets perform trajectory simulations using an 8-cycle and an ultra-short single-cycle linearly polarized near-infrared (NIR) laser pulse, respectively. The resulting PMDs are displayed in Figs. 3 and 4. To be faithful to the original work (Ref. [45]) where the ADK initial condition was used with no prefactor included ($\mathcal{P} = 1$), we used the ADK-QTMC and ADK-SCTS schemes and set `rate_prefix=:Exp`. Comparison between the PMD obtained with the QTMC and SCTS phase methods reveals underestimation of the Coulomb interaction's influence on the phase by the QTMC method, which is observed in the number of angular nodes in the low-energy structures.

Furthermore, the trajectory-based methods exhibit certain limitations, as evidenced by discrepancies between the PMD structures obtained from the ADK-SCTS and the TDSE: Firstly, the trajectory-based methods have limited capability in reproducing the high-energy signals in the PMD, which are predominantly associated with recollision processes; Secondly, despite the improvements in the quantum phase expressions introduced by the SCTS method, minor discrepancies persist in the low-energy structures of the PMD when compared to those generated by the TDSE. These observations address the current limitations of trajectory-based methods.

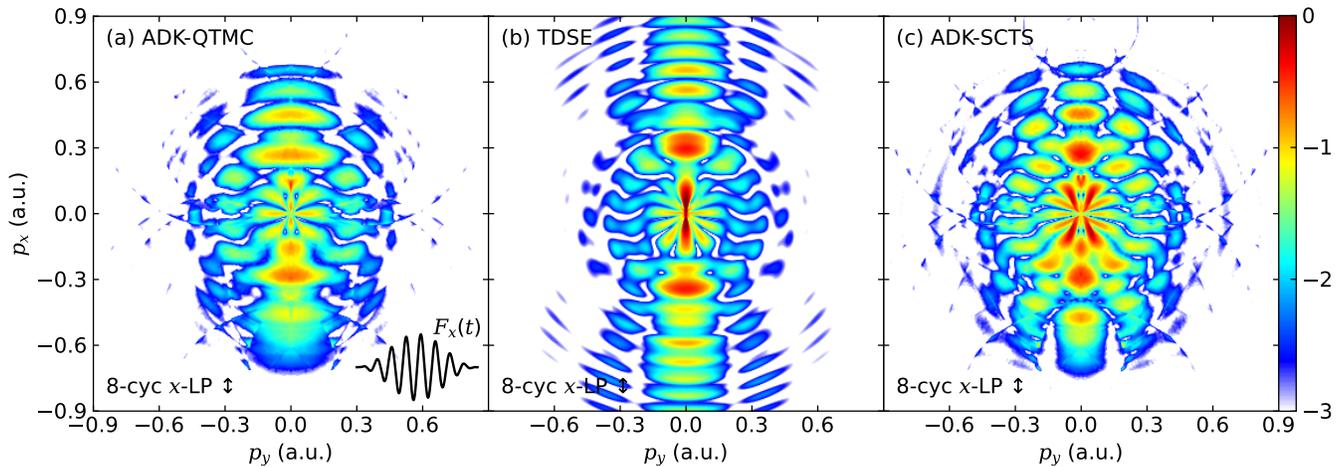

FIG. 3. Low-energy structure of the PMD from interaction of a hydrogen atom with an 8-cycle linearly-polarized laser pulse obtained using (a) the QTMC and (c) the SCTS phase methods, as well as (b) the TDSE simulation for reference (logarithmic scale, each normalized to the maximum value). The laser, polarized along the $x$-axis, has a wavelength of 800 nm and a peak intensity of 90 TW/cm². The temporal profile of the electric field strength is illustrated in the left sub-figure.

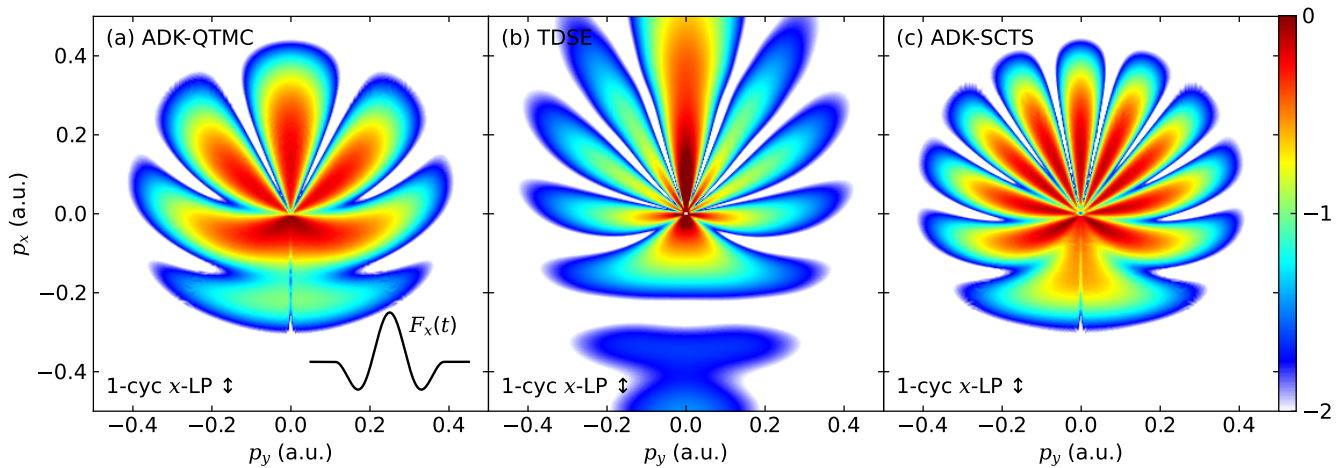

FIG. 4. Low-energy structure of the PMD from the interaction of a hydrogen atom with a single-cycle linearly-polarized laser pulse obtained using (a) the QTMC and (c) the SCTS phase methods, as well as (b) the TDSE simulation for comparison (logarithmic scale, each normalized to the maximum value). The laser is an 800-nm single-cycle pulse polarized along the $x$-axis, with a peak intensity of 90 TW/cm². The CEP of the laser is set to $\pi/2$ to achieve the desired pulse shape, as the PMD is highly sensitive to the CEP for ultra-short pulses. The temporal profile of the electric field strength is depicted in the left sub-figure.

```
examples/test_8cycs_LP.jl

1  using eTraj
2  using eTraj.Targets, eTraj.Lasers, eTraj.Units
3
4  l = Cos2Laser(peak_int=90.0TW/cm^2, wave_len=800.0nm, cyc_num=8, ellip=0.0)
5  t = get_atom("H")
6
7  for phase_method in [:QTMC, :SCTS]
8      perform_traj_simulation(
9          init_cond_method    = :ADK,
10         rate_prefix         = :Exp,
```

```
11        traj_phase_method   = phase_method,
12        laser               = l,
13        target              = t,
14        dimension           = 2,
15        sample_t_intv       = (-350,350),
16        sample_t_num        = 50000,
17        traj_t_final        = 500,
18        final_p_max         = (1.0,1.0),
19        final_p_num         = (500,500),
20        ss_kd_max           = 1.0,
21        ss_kd_num           = 20000,
22        output_path         = "ADK-$(phase_method)_9e13_800nm_8cyc_LP_ExpRate.jld2"
23    )
24 end
```

―――――――――――――――――――― examples/test_1cyc_LP.jl ――――――――――――――――――――

```
1  using eTraj
2  using eTraj.Targets, eTraj.Lasers, eTraj.Units
3
4  l = Cos2Laser(peak_int=90.0TW/cm^2, wave_len=800.0nm, cyc_num=1, cep=π/2, ellip=0.0)
5  t = get_atom("H"; soft_core=1e-12)
6
7  for phase_method in [:QTMC, :SCTS]
8      perform_traj_simulation(
9          init_cond_method    = :ADK,
10         rate_prefix         = :Exp,
11         traj_phase_method   = phase_method,
12         laser               = l,
13         target              = t,
14         dimension           = 2,
15         sample_t_intv       = (-50,50),
16         sample_t_num        = 30000,
17         traj_t_final        = 100,
18         final_p_max         = (1.0,1.0),
19         final_p_num         = (500,500),
20         ss_kd_max           = 1.5,
21         ss_kd_num           = 10000,
22         output_path         = "ADK-$(phase_method)_9e13_800nm_1cyc_LP_ExpRate.jld2"
23     )
24 end
```

### C.  Interaction with an $\omega - 2\omega$ Bichromatic Clover-shaped Laser

Bichromatic laser fields, comprising a fundamental frequency component and its second harmonic, enable the construction of tailored waveforms that facilitate precise control and investigation of ultrafast light-matter interaction dynamics [92–96].

In this example, we use a bichromatic laser pulse which combines two counter-rotating circularly-polarized laser pulses of 800 nm and 400 nm wavelengths. By adjusting the relative intensity of the two frequency components, the waveform of the pulse can be tailored to exhibit a clover-like shape (see Fig. 5), which facilitates control of the emission direction of the ejected electrons. The PMDs for different laser intensities are shown in Fig. 6.

```
                           examples/test_Bichromatic_CCP.jl
1   using eTraj
2   using eTraj.Targets, eTraj.Lasers, eTraj.Units
3
4   for int in [1e14, 3e14, 5e14, 7e14]
5       @info "Running I0=$(int) W/cm^2"
6       l1 = Cos2Laser(peak_int=int*W/cm^2, wave_len=800.0nm, cyc_num=8,  ellip= 1.0)
7       l2 = Cos2Laser(peak_int=int*W/cm^2, wave_len=400.0nm, cyc_num=16, ellip=-1.0)
8       l = BichromaticLaser(l1=l1, l2=l2)
9       t = get_atom("H")
10      perform_traj_simulation(
11          init_cond_method    = :ADK,
12          rate_prefix         = Set([:Pre,:Jac]),
13          traj_phase_method   = :SCTS,
14          laser               = l,
15          target              = t,
16          dimension           = 2,
17          sample_t_intv       = (-350,350),
18          sample_t_num        = 10000,
19          traj_t_final        = 450,
20          final_p_max         = (2.5,2.5),
21          final_p_num         = (500,500),
22          ss_kd_max           = 1.0,
23          ss_kd_num           = 5000,
24          output_path         = "ADK-SCTS_Bichromatic_$(int)_800+400nm_8+16cycs_CounterCP.jld2"
25      )
26  end
```

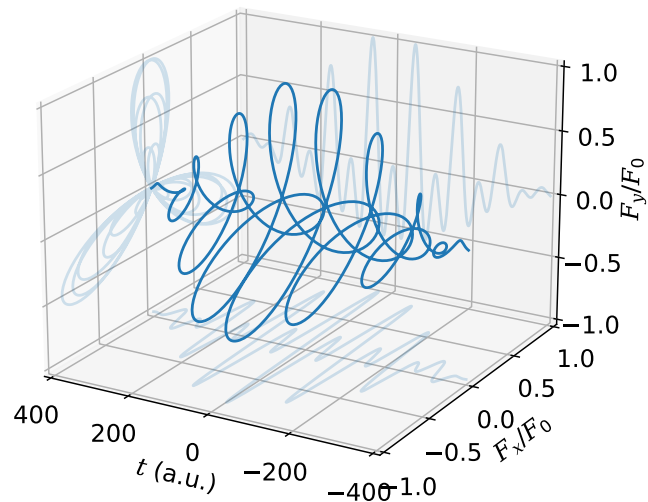

FIG. 5.   Waveform of the electric field of an $\omega - 2\omega$ bichromatic counter-rotating circularly-polarized laser pulse. The fundamental (800 nm) and the second harmonic (400 nm) components of the laser share the same peak intensity and duration (8 cycles of the fundamental pulse).

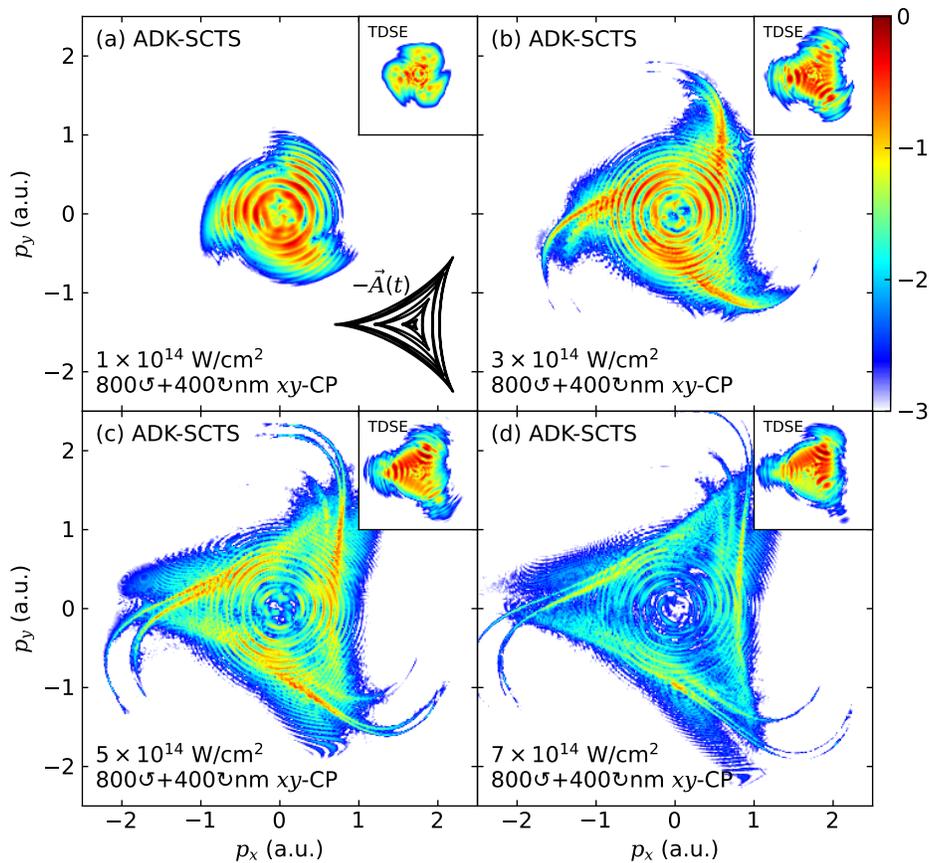

FIG. 6. PMDs of the interaction of a hydrogen atom with an $\omega - 2\omega$ bichromatic counter-rotating circularly-polarized laser pulse of different peak intensities (logarithmic scale, each normalized to the maximum value). PMDs obtained by TDSE simulation are shown in the top-right corner of each sub-figure for comparison. The laser pulse's waveform of electric field and parameters are illustrated in Fig. 5, and the trace of $-\vec{A}(t)$ is displayed in the first sub-figure.

## D.  WFAT-CTMC Simulation of Molecular Targets

The WFAT framework offers a rigorous approach for computing tunneling ionization probabilities in molecular systems, particularly advantageous for complex molecular structures. This section demonstrates the application of the WFAT-CTMC simulation scheme to various molecular targets.

─── **examples/test_Molecules.jl** ───

```
using eTraj
using eTraj.Targets, eTraj.Lasers, eTraj.Units

l = Cos2Laser(peak_int=4e14W/cm^2, wave_len=800.0nm, cyc_num=6, ellip=1.0)
t = [get_mol("Hydrogen"; rot_β=90°),
     get_mol("Carbon Monoxide"; rot_β=90°),
     get_mol("Oxygen"; rot_β=90°),
     get_mol("Oxygen"; rot_β=90°),
     get_mol("Benzene"; rot_β=90°),
     get_mol("Benzene"; rot_β=90°)]
orbit_ridx = [0, 0, (1,0), (1,-1), 0, -1]
path = [
    "WFAT-CTMC_Hydrogen_HOMO_4e14_800nm_6cyc_CP.jld2",
    "WFAT-CTMC_CarbonMonoxide_HOMO_4e14_800nm_6cyc_CP.jld2",
```

```
15          "WFAT-CTMC_Oxygen_α-HOMO_4e14_800nm_6cyc_CP.jld2",
16          "WFAT-CTMC_Oxygen_α-HOMO-1_4e14_800nm_6cyc_CP.jld2",
17          "WFAT-CTMC_Benzene_HOMO_4e14_800nm_6cyc_CP.jld2",
18          "WFAT-CTMC_Benzene_HOMO-1_4e14_800nm_6cyc_CP.jld2"
19      ]
20  for i in eachindex(t)
21      perform_traj_simulation(
22          init_cond_method    = :WFAT,
23          traj_phase_method   = :CTMC,      # WFAT supports CTMC only
24          laser               = l,
25          target              = t[i],
26          mol_orbit_ridx      = orbit_ridx[i],
27          dimension           = 2,
28          sample_t_intv       = (-300,300),
29          sample_t_num        = 10000,
30          traj_t_final        = 350,
31          final_p_max         = (2.0,2.0),
32          final_p_num         = (500,500),
33          ss_kd_max           = 2.0,
34          ss_kd_num           = 5000,
35          output_path         = path[i]
36      )
37  end
```

The simulation employs a 6-cycle, circularly polarized 800-nm laser pulse. This duration ensures PMD symmetry for isotropic atoms while enabling effective orbital imaging through the PMD structure: nodal planes in the molecular orbital manifest as nodes or dark regions in the PMD when intersected by the laser's electric field vector. The molecular orientation is configured to align the molecular frame (MF) $z$-axis parallel to the laboratory frame (LF) $x$-axis, achieved through a 90° counterclockwise rotation around the $y$-axis (specified by `rot_β=90°`).

The PMDs obtained from the HOMOs of different molecules are presented in Fig. 7. The structures of the PMDs reflect the geometries of the corresponding orbitals, which are further mirrored in the orientation-dependent structure factors, as illustrated in Fig. 8.

Fig. 7 (a) displays the PMD of the hydrogen molecule's HOMO orbital, $1s\sigma_g$, which primarily arises from the in-phase combination of the two 1s orbitals of the hydrogen atoms. The overall spherical shape of the orbital is consistent with the lowest-order $[\nu = (n_\xi, m) = (0, 0)]$ squared structure factor $|G_{00}|^2$ shown in Fig. 8 (a), resulting in the evenly distributed ring-like structure of the PMD.

The carbon monoxide (CO) molecule, a heteronuclear diatomic molecule, has a HOMO orbital designated as $3\sigma_g$. The $3\sigma_g$ orbital of the CO molecule differs from a conventional $\sigma_g$ orbital of a homonuclear molecule, as the electron density is predominantly localized on the carbon atom [see Fig. 8 (b)]. This localization leads to a significant increase in the squared structure factor and ionization probability when the negative electric field is oriented toward the carbon atom. This localized peak within the ring structure is evident in the PMD of the CO molecule, as shown in Fig. 7 (b).

The HOMOs of the oxygen ($O_2$) and benzene ($C_6H_6$) molecules, depicted in Figs. 8 (c) and (d), respectively, are degenerate orbitals with $\pi$ symmetry. The oxygen molecule's 'α-HOMO' (one of the $2p\pi_u$ orbitals) and the benzene molecule's 'HOMO' ($2p\pi_3$), after a rotation of `rot_β=90°`, exhibit nodal planes along the $x-z$ and $y-z$ directions, respectively. Consequently, the rotating electric field in the $x-y$ plane interacts with a four-lobe structure, as revealed in the PMDs shown in Figs. 7 (c1) and (d1). In contrast, the oxygen molecule's 'α-HOMO-1' (another $2p\pi_u$ orbital) and the benzene molecule's 'HOMO-1' ($2p\pi_2$) each possess a nodal plane in the $x-y$ plane. This implies that, in the zeroth order ($m = 0$), the outgoing electron waves originating from the '+' and '−' regions of the orbital interfere destructively, resulting in a net-zero ionization probability. For non-zero-$m$ channels, $\mathcal{W}_\nu(F, k_t = 0) \equiv 0$ [see Eq. (58)], leading to the appearance of a nodal ring corresponding to zero initial momenta ($k_t = 0$) in the PMDs of Figs. 7 (c2) and (d2). Under these conditions, ionization is suppressed, and the first-order channels ($m = \pm 1$) dominate the contribution to the ionization probability.

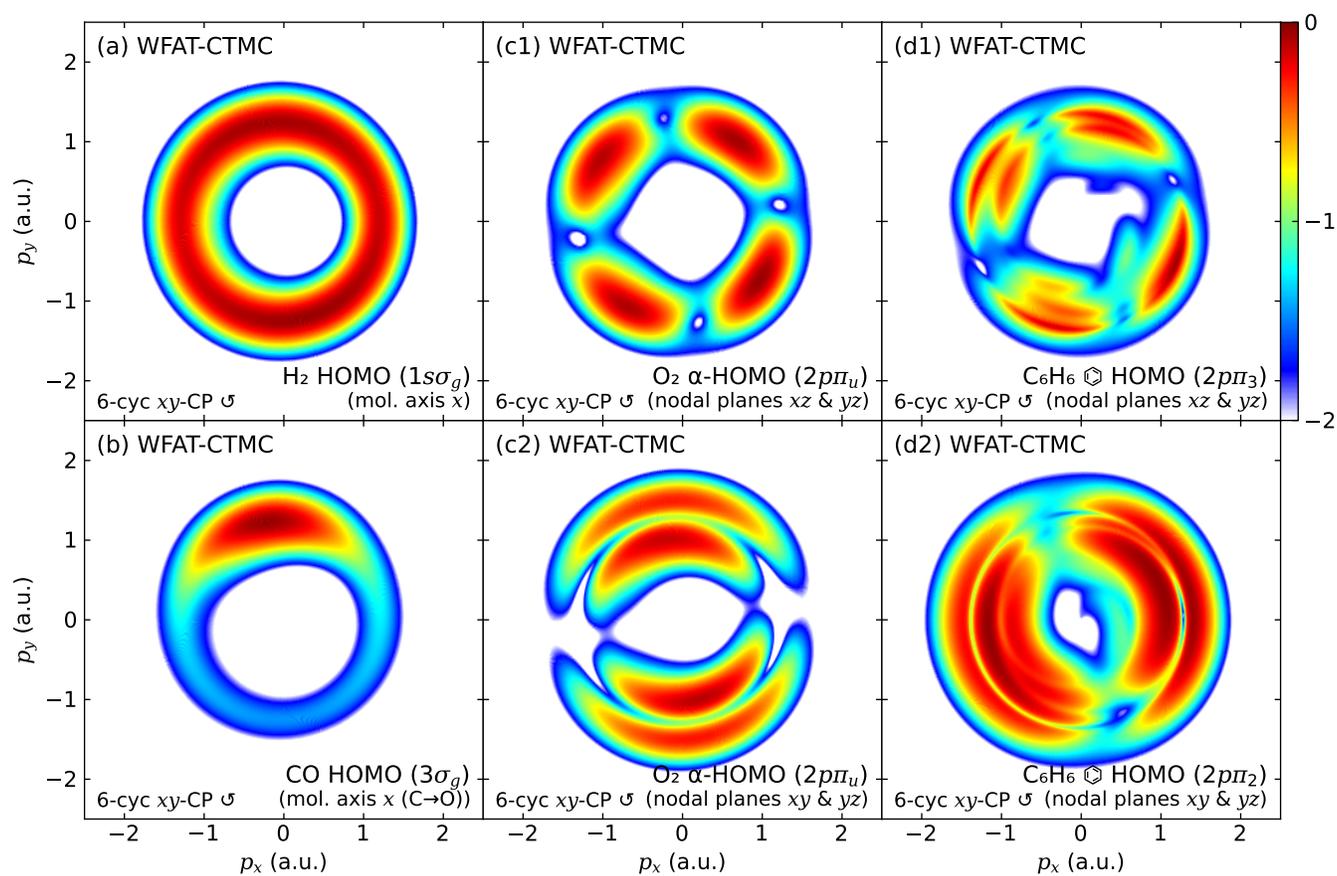

FIG. 7. PMDs of the interaction of different molecules' HOMOs with a 6-cycle circularly-polarized laser pulse obtained by the WFAT-CTMC scheme (logarithmic scale, each normalized to the maximum value). The laser's peak intensity is 0.4 PW/cm² and the wavelength is 800 nm.

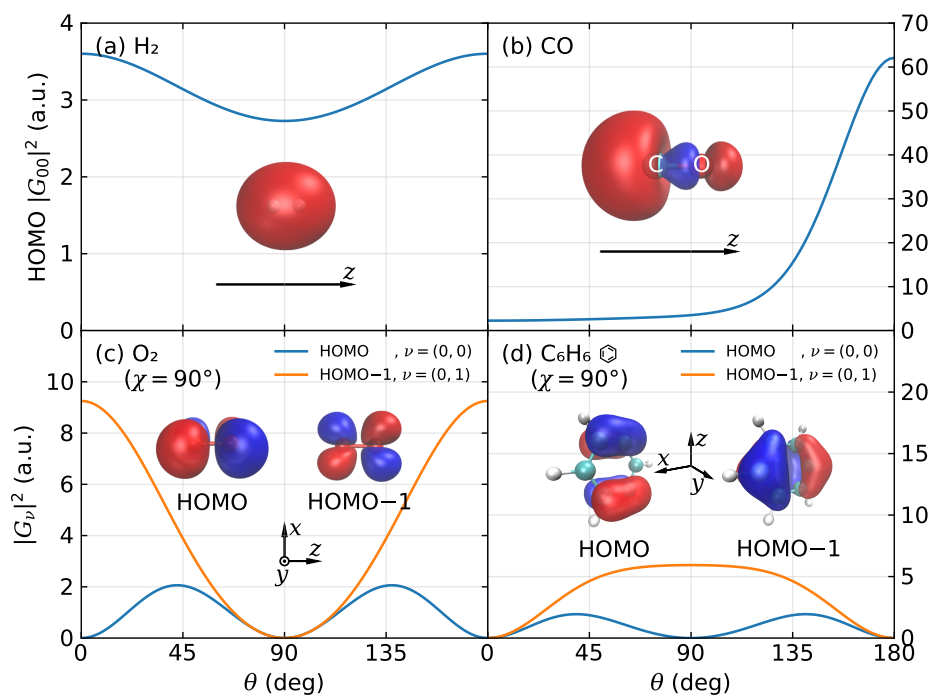

FIG. 8. Squared structure factors $|G_\nu(\theta, \chi = 90°)|^2$ and the wavefunction isosurface diagram of the different molecules' HOMOs (in MF). The 'HOMO' and 'HOMO−1' orbitals are actually degenerate and are only used to distinguish between each other in the program. Wavefunction diagrams are created using the VMD [97].

## V. CONCLUSIONS AND PROSPECTS

The classical/semiclassical trajectory-based methods have long been recognized as powerful tools for simulating and understanding strong-field phenomena. In this article, we have presented a comprehensive and unified theoretical framework for these methods, along with the `eTraj` program package, which enables researchers to perform trajectory simulations to determine the photoelectron momentum distribution resulting from strong-field ionization of atoms and molecules. The `eTraj` package is designed to be versatile, efficient, flexible, and user-friendly, making it an accessible tool for the strong-field physics community. Key features of `eTraj` include:

- **Unified Theoretical Framework**: `eTraj` integrates a variety of established initial condition methods, including the Strong-Field Approximation with Saddle-Point Approximation (SFA-SPA), SFA-SPA with Non-adiabatic Expansion (SFA-SPANE), Ammosov-Delone-Krainov (ADK) theory, and Weak-Field Asymptotic Theory (WFAT) for molecules. It is important to note that this unified framework is not merely a collection of formulas from earlier literature. Instead, we have developed a cohesive approach that places these methods, each with different levels of non-adiabaticity, on the same footing. This ensures that they are directly comparable in terms of absolute ionization rates. This unified framework allows users to choose the most suitable method for their specific research needs, whether they are studying atomic or molecular systems.

- **Efficient and Versatile Simulation**: The program supports multiple trajectory evolution methods, such as Classical Trajectory Monte-Carlo (CTMC), Quantum Trajectory Monte-Carlo (QTMC), and Semiclassical Two-Step (SCTS) model. These methods provide different levels of quantum phase treatment, enabling users to balance computational efficiency with the inclusion of quantum effects. The versatility of `eTraj` is further enhanced by its compatibility with various laser types and arbitrary molecular configurations, making it suitable for a wide range of experimental setups.

- **High Flexibility**: The `eTraj` package is highly flexible, supporting a wide range of targets, including both simple and complex atoms and molecules, neutral and charged species, and systems in either ground or excited states; this feature also extends to laser parameter settings, allowing for customizable control over the temporal and polarization profiles of laser fields as well as the definition of bichromatic fields. This flexibility ensures that `eTraj` can be applied to a broad spectrum of research problems in strong-field physics, from fundamental studies of atomic ionization to complex molecular dynamics.

- **User-Friendly Design**: `eTraj` is designed with user-friendliness in mind. It features clear documentation, intuitive syntax, and built-in support for parallel computing, allowing researchers to focus on their scientific questions rather than the intricacies of the computational implementation. The program is also compatible with multiple operating systems, including Linux, macOS, and Windows (with some limitations for molecular calculations on Windows).

- **Future Extensions**: Beyond its current capabilities, `eTraj` is poised for future development. We plan to extend the program to support trajectory simulations beyond the dipole approximation, which will enable the study of light momentum transfer when interacting with matter. We also plan to expand the laser geometry capabilities of `eTraj` to support non-collinear laser configurations, allowing researchers to explore a wider range of experimental setups and more complex light-matter interactions. Additionally, we aim to introduce features for tracking selected electron trajectories, providing deeper insights into the dynamics of individual ionization events. The `eTraj` package can be further extended to support two-electron simulations, albeit limited to purely classical simulations starting from a microcanonical ensemble of electrons. This extension will facilitate the study of electron-correlation mediated processes, such as non-sequential double ionization.

In summary, `eTraj` represents a significant advancement in the simulation of strong-field ionization processes. Its unified theoretical framework, efficient and versatile design, flexible nature, and user-friendly implementation make it a powerful tool for researchers in the field of strong-field physics. We anticipate that `eTraj` will become an indispensable resource, enabling the community to push the boundaries of our understanding of light-matter interactions and explore new frontiers in attosecond and ultrafast science.

## Appendix A: Ionization Rate of SFA-SPA in the Adiabatic Limit and the Coulomb-correction Term

In the adiabatic limit, the transition amplitude of the SFA-SPA, given by Eq. (33), can be expanded to yield a more explicit expression.

We begin with the spherical harmonics $Y_{lm}[\hat{\boldsymbol{k}}(t_s)]$ in Eq. (34), which is expressed as

$$Y_{lm}(\theta, \phi) = (-1)^m \sqrt{\frac{2l+1}{2} \frac{(l-m)!}{(l+m)!}} P_l^m(\cos\theta) \frac{\mathrm{e}^{im\phi}}{\sqrt{2\pi}}, \tag{A1}$$

where $P_l^m$ denotes the associated Legendre polynomial, and the $z$-axis is aligned with the direction of the field polarization. Although $\hat{\boldsymbol{k}}(t_s)$ is a complex vector, $\cos\theta$ adheres to the conventional definition, as given by Eq. (30) in Ref. [64]:

$$\cos\theta = \frac{\boldsymbol{k}(t_s) \cdot \boldsymbol{F}(t_r)}{\sqrt{\boldsymbol{k}(t_s) \cdot \boldsymbol{k}(t_s)} F(t_r)} = \sqrt{1 + \frac{k_t^2}{\kappa^2}}. \tag{A2}$$

By substituting Eq. (A2) into Eq. (A1) and employing the connection formula for $P_l^m(x)$ (Eq. (14.9.3) of Ref. [98]):

$$P_l^{-m}(x) = (-1)^m \frac{(l-m)!}{(l+m)!} P_l^m(x), \tag{A3}$$

along with the asymptotic behavior of $P_l^m(x)$ as $x \to 1^+$ (Eqs. (14.8.7) and (14.8.8) of Ref. [98]):

$$P_l^m(x) \sim \frac{1}{|m|!} \left(\frac{x-1}{2}\right)^{|m|/2} \times \begin{cases} (l+m)!/(l-m)!, & \text{for } m = 0, 1, 2, \cdots, \\ 1, & \text{for } m = -1, -2, \cdots, \end{cases} \tag{A4}$$

we derive the leading-order term of $Y_{lm}[\hat{\boldsymbol{k}}(t_s)]$ for small $k_t$:

$$Y_{lm}[\hat{\boldsymbol{k}}(t_s)] \sim Q_{lm} \frac{(k_t/\kappa)^{|m|}}{2^{|m|}|m|!} \frac{\mathrm{e}^{im\phi}}{\sqrt{2\pi}}, \tag{A5}$$

where the coefficient is given by

$$Q_{lm} = (-1)^m \sqrt{\frac{2l+1}{2} \frac{(l+|m|)!}{(l-|m|)!}}. \tag{A6}$$

Next, we address the Jacobian factor $J(t_r, k_\perp)$ [Eq. (25)]. In the adiabatic limit, the Jacobian simplifies to:

$$J(t_r, k_\perp) = \left|\frac{\partial(p_x, p_y)}{\partial(t_r, k_\perp)}\right| = F(t_r) + \frac{k_\perp}{F^2(t_r)}\left[F_x(t_r)F_y'(t_r) - F_x'(t_r)F_y(t_r)\right] = F(t_r), \tag{A7}$$

where the second equality holds for $k_\parallel = 0$, and the third equality is valid under the adiabatic condition $\boldsymbol{F}'(t_r) = 0$. Here, $p_x = k_\perp F_y(t_r)/F(t_r) - A_x(t_r)$ and $p_y = -k_\perp F_x(t_r)/F(t_r) - A_y(t_r)$.

Combining the above results, we obtain

$$\begin{aligned}
\frac{\mathrm{d}W}{\mathrm{d}t_r \mathrm{d}\boldsymbol{k}_t} &\approx J(t_r, k_\perp) |\mathcal{P}_{\boldsymbol{p}}^{\mathrm{ADK}}(t_s)|^2 \exp\left[-\frac{2}{3}\frac{(k_t^2 + \kappa^2)^{3/2}}{F}\right] \\
&= \frac{C_{\kappa l}^2}{\pi}\left[\frac{\Gamma(n^*/2+1)}{|m|!}\right]^2 |Q_{lm}|^2 2^{n^*-2|m|+1} F^{-n^*} \kappa^{4n^*-2|m|+1} k_t^{2|m|} (\kappa^2 + k_t^2)^{-(n^*+1)/2} \exp\left[-\frac{2}{3}\frac{(k_t^2 + \kappa^2)^{3/2}}{F}\right].
\end{aligned} \tag{A8}$$

Integrating Eq. (A8) over $\boldsymbol{k}_t$ in polar coordinates under the small-$k_t$ approximation yields:

$$\begin{aligned}
\int \mathrm{d}\boldsymbol{k}_t k_t^{2|m|} (\kappa^2 + k_t^2)^{-(n^*+1)/2} &\exp\left[-\frac{2}{3}\frac{(k_t^2 + \kappa^2)^{3/2}}{F}\right] \\
&\approx 2\pi\kappa^{-(n^*+1)} \exp\left(-\frac{2\kappa^3}{3F}\right) \int_0^\infty \mathrm{d}k_t \, k_t^{2|m|+1} \exp\left(-\frac{\kappa k_t^2}{F}\right) \\
&= \pi F^{|m|+1} \kappa^{-n^*-|m|-2} |m|! \exp\left(-\frac{2\kappa^3}{3F}\right),
\end{aligned} \tag{A9}$$

leading to

$$w = \frac{dW}{dt_r} = C_{\kappa l}^2 \left[ \Gamma \left( \frac{n^*}{2} + 1 \right) \right]^2 \frac{|Q_{lm}|^2}{|m|!} 2^{n^*-2|m|+1} F^{-n^*+|m|+1} \kappa^{3n^*-3|m|-1} \exp \left( -\frac{2\kappa^3}{3F} \right).$$ (A10)

Eq. (A10) resembles the ADK ionization rate [99, 100], but differs in the power coefficients of the factors 2, $F$, and $\kappa$. The discrepancy arises because our derivation neglects the influence of the Coulomb potential on the phase $S_{\boldsymbol{p}}$. In our theoretical formulation, this effect is only partially included through the use of the asymptotic wavefunction in the Coulomb potential, as described in Eq. (12).

In Refs. [101–103], an alternative approach employing a trajectory-based perturbation method was used to study the influence of the Coulomb interaction, yielding the Coulomb-correction (CC) term:

$$\delta S_{\boldsymbol{p}}(t_s) = -in^* \ln \frac{2\kappa^3}{F(t_r)},$$ (A11)

which corrects the ionization rate for the *short-range* (SR) case as follows:

$$w^{\text{Coulomb}} = \exp(-2\Im \delta S_{\boldsymbol{p}}) w^{\text{SR}} = \left( \frac{2\kappa^3}{F} \right)^{2n^*} w^{\text{SR}}.$$ (A12)

To apply the Coulomb-correction term to Eq. (A10), we first remove the partial Coulomb correction by setting $n^* = 0$, yielding the short-range ionization rate $w^{\text{SR}}$. Applying the correction in Eq. (A12), we recover the ionization rate expression, which matches the ADK rate:

$$w^{\text{ADK}} = \frac{C_{\kappa l}^2 |Q_{lm}|^2}{|m|!} 2^{2n^*-2|m|+1} F^{-2n^*+|m|+1} \kappa^{6n^*-3|m|-1} \exp \left( -\frac{2\kappa^3}{3F} \right).$$ (A13)

This suggests that an additional Coulomb-correction factor applied to our ionization probability in Sec. II A, based on the SFA, could improve alignment with experimental results:

$$C^{\text{CC,ADK}} = \left( \frac{2\kappa^3}{F} \right)^{n^*} \left[ \Gamma \left( \frac{n^*}{2} + 1 \right) \right]^{-2}.$$ (A14)

Ref. [103] provides a correction term applicable for arbitrary Keldysh parameters:

$$C^{\text{CC}} = \left( \frac{2\kappa^3}{F} \right)^{n^*} (1 + 2\gamma/e)^{-n^*} \left[ \Gamma \left( \frac{n^*}{2} + 1 \right) \right]^{-2},$$ (A15)

which has been incorporated into `eTraj`.

## ACKNOWLEDGMENTS

We would like to thank Nikolay I. Shvetsov-Shilovski, Emilio Pisanty, Simon Brennecke, Xiaodan Mao and Kefei Wu as well as other researchers for their helpful discussions. This work was supported by the National Natural Science Foundation of China (Grant Nos. 92150105, 12474341, 12227807, and 12241407), the Science and Technology Commission of Shanghai Municipality (Grant No. 23JC1402000), and the Shanghai Pilot Program for Basic Research (Grant No. TQ20240204). Numerical computations were in part performed on the ECNU Multifunctional Platform for Innovation (001).

---